\documentclass[aps,prl,twocolumn,superscriptaddress]{revtex4-1}
\usepackage{graphicx}
\usepackage{dcolumn}
\usepackage{bm}
\usepackage{amsmath}
\usepackage{amssymb}
\usepackage{latexsym}
\usepackage{epsfig}
\usepackage{amsbsy}
\usepackage{array}
\usepackage{amssymb}
\usepackage{setspace}
\usepackage{float,url}
\usepackage{textcomp}
\usepackage{multirow}
\usepackage{threeparttable}
\usepackage[colorlinks=true,
                        linkcolor=blue,
                        anchorcolor=blue,
                     urlcolor=blue,
                        citecolor=blue]{hyperref}

\begin{document}

\title{Nuclear Zeeman Effect on Heading Errors and the Suppression in Atomic
Magnetometers}
\author{Yue Chang}
\email{yuechang7@gmail.com}
\author{Yu-Hao Guo}
\author{Shuang-Ai Wan}
\author{Jie Qin}
\email{jie.qin@yahoo.com}

\begin{abstract}
Precession frequencies measured by optically-pumped scalar magnetometers are
dependent on the relative angle between the sensor and the external magnetic
field, resulting in the so-called heading errors if the magnetic field
orientation is not well known or is not stable. The heading error has been
known to be caused mainly by the nonlinear Zeeman effect and the
orientation-dependent light shift. In this work, we find that the nuclear
Zeeman effect can also a significant impact on the heading errors,
especially for continuously-driving magnetometers with unresolved magnetic
transitions. It not only shifts the precession frequency but deforms the
heading errors and causes asymmetry: the heading errors for pump lasers with
opposite helicities are different. The heading error also depends on the
relative direction (parallel or vertical) of the probe laser to the RF
driving magnetic field. Thus, one can design the configuration of the
magnetometer and make it work in the smaller-heading-error regime. To
suppress the heading error, our studies suggest to sum up the output
precession frequencies from atomic cells pumped by two lasers with opposite
helicities and probed by lasers propagating in orthogonal directions (one
parallel and another perpendicular to the RF field), instead of utilizing
probe lasers propagating in the same directions. Due to the nuclear Zeeman
effect, the average precession frequencies in the latter case can have a
non-negligible angular dependence, while in the former case the
nuclear-Zeeman-effect induced heading error can be largely compensated and
the residue is within $1 $Hz. Furthermore, for practical use, we propose to
simply utilize a small magnetic field parallel/antiparallel to the pump
laser. By tuning the magnitude of this auxiliary field, the heading error
can be flattened around different angles, which can improve the accuracy
when the magnetometer works around a certain orientation angle.
\end{abstract}

\maketitle

\affiliation{Beijing Automation Control Equipment Institute, Beijing 100074,
China}
\affiliation{Quantum Technology R$\&$D Center of China Aerospace
Science and Industry Corporation, Beijing 100074, China}

\flushbottom

Optically-pumped magnetometers have achieved high sensitivity \cite%
{Kominis2003,Budker2003,Budker2007} and have been applied in a broad range
from archaeology and geophysics \cite%
{kvamme2006magnetometry,gaffney2008detecting,dang2010ultrahigh,PhysRevLett.103.261801}
to fundamental physics \cite%
{fortson2003search,PhysRevA.75.063416,roberts2015}. In scalar atomic
magnetometers, the magnitude of the external magnetic field is determined by
measuring the precession frequency of alkali-metal atoms. This frequency is
dependent on the sensor's orientation with respect to the magnetic field,
resulting in the so-called heading errors \cite%
{doi:10.1063/1.1663412,alexandrov2003recent,PhysRevLett.105.193601,PhysRevLett.120.033202}%
, which is one of the major sources of accuracy degradation especially for
magnetometers operating in the geophysical range ($20-80$~$\mu $T).

\begin{figure}[tb]
\begin{center}
\includegraphics[width=\linewidth]{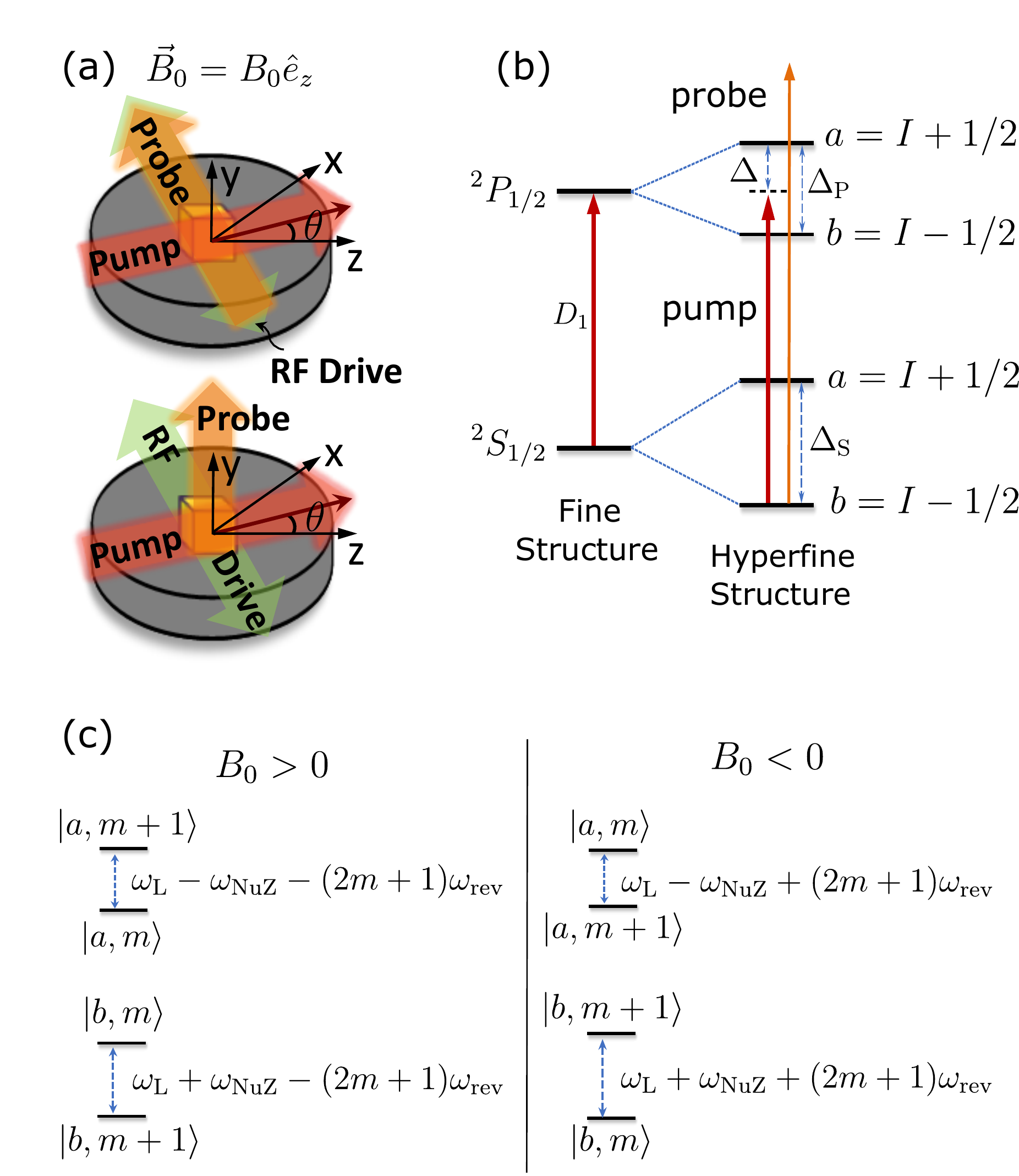}
\end{center}
\caption{(a) Schematic of an atomic magnetometer fixed on a rotatable table
(the gray disc). Here, an atomic cell (yellow cubic at the center)
containing alkali-metal atoms and buffer gas is pumped by a
circularly-polarized laser (red arrow) propagating in the $xz$ plane with a
tilted angle $\protect\theta $ to the external field $\vec{B}_{0}=B_{0}\hat{e%
}_{z}$. A small RF magnetic field (green double arrow) drives the atoms and
the induced precession is measured by the linearly-polarized probe laser
(orange arrow) propagating parallel (upper figure) or perpendicular (lower
figure) to the RF field. (b) Fine states and hyperfine states of the
alkali-metal atom. The pump laser induces transitions (D1 transition)
between the ground states $\left\vert Fm\right\rangle _{S}$ and the first
excited states $\left\vert Fm\right\rangle _{P}$ with $F=a,b$, and $m$ the
magnetic number. The probe laser is far detuned from this D1 transition. (c)
Energy level spacing between two adjacent ground-state Zeeman sublevels of
the alkali-metal atom. Besides the Larmor frequency$\ \protect\omega _{%
\mathrm{L}}\equiv \protect\mu _{\mathrm{eff}}\left\vert B_{0}\right\vert $
and the quantum beat revival frequency $\protect\omega _{\mathrm{rev}}\equiv
\protect\mu _{\mathrm{eff}}^{2}B_{0}^{2}/\Delta _{S}$, there is the third
term $\protect\omega _{\mathrm{NuZ}}\equiv g_{I}\protect\mu _{\mathrm{N}%
}\left\vert B_{0}\right\vert $ coming from the linear NuZ splitting, leading
to different precession frequencies in the $a$ and $b$ hyperfine manifolds.}
\label{fig1}
\end{figure}

Heading errors in atomic magnetometers have been studied both theoretically
and experimentally \cite%
{doi:10.1063/1.1663412,Hovde2011HeadingEI,2017ApPhB.123...35C,PhysRevLett.120.033202,PhysRevA.99.013420}%
. It has been shown \cite{PhysRevA.99.013420} that the main contributions to
heading errors in continuously-pumped magnetometers are the nonlinear Zeeman
(NLZ) effects \cite%
{PhysRevA.75.051407,PhysRevLett.120.033202,PhysRevA.79.023406,PhysRevA.82.023417}
and the light shift (LS) \cite%
{RevModPhys.44.169,scholtes2012light,PhysRevA.58.1412,PhysRevA.99.063411}
(in continuously-driving systems) that is orientation dependent. In most of
these studies, apart from a small correction to the Larmor frequency, the
interaction between the external magnetic field and the nuclear spins is
neglected since the nuclear magneton $\mu _{\mathrm{N}}$ is about $3$ orders
smaller than the Bohr magneton $\mu _{\mathrm{B}}$. However, in the
geophysical field range, the linear nuclear Zeeman (NuZ) splitting is larger
than or comparable with the NLZ shift. For instance, for $\left. ^{85}\text{%
Rb}\right. $ in a field $B_{0}=55\mathrm{\mu T}$, the former is about $226$%
Hz while the latter is about $22$Hz. When the frequencies of transitions
between adjacent magnetic levels are not resolved, for instance, when the
linewidth of the individual transitions is relatively large in the presence
of several hundreds of buffer gas, the NuZ effect is non-negligible for the
heading errors, especially for continuously-driving magnetometers. In this
paper, we theoretically and experimentally study the heading error in atomic
magnetometers by including the NuZ effect and find that it can significantly
modify the heading errors. The setup for our study is schematically shown in
Fig.~\ref{fig1}(a), where an atomic cell containing alkali-metal atoms and
buffer gas ($\mathrm{N_{2}}$) is exposed to the external magnetic field $%
\vec{B}_{0}=B_{0}\hat{e}_{z}$ along the $z$ direction. The
circularly-polarized pump laser, whose propagation direction together with $%
\vec{B}_{0}$ defines the $xz$ plane, is tilted by an angle $\theta $ to $%
\vec{B}_{0}$. An oscillating magnetic field perpendicular to the pump laser
is generated by RF coils to induce atomic spin polarizations in the $xy$
plane, which is reconstructed by measuring the optical rotation of a
linearly-polarized probe laser. Without loss of generality, we assume the RF
field is in the $xz$ plane and the probe laser propagates parallel or
perpendicular (along the $y$ direction) to the RF field since only their
relative direction matters. For parallel probe lasers, we find that,
compared to the case without the NuZ effect, the heading error is
smaller/larger when the pump laser is left/right-handed circularly polarized
($\sigma ^{+}/\sigma ^{-}$). For vertical probe lasers, however, it is
another way around. This suggests to reduce the heading errors by employing
two pump lasers with opposite helicities and two probe lasers with one
parallel while another perpendicular to the driving field. Furthermore, for
practical use, we propose a simple scheme to suppress the heading error by
utilizing a small magnetic field parallel (for $\sigma ^{+}$ polarization)
or antiparallel (for $\sigma ^{-}$ polarization) to the pump laser. By
tuning the magnitude of this auxiliary field, the heading error can be
flattened around different orientation angles.

It can be proved \cite{supmat} that the precession frequency is invariant
when changing $\theta $ to$\ -\theta $, and inverting the helicity of the
pump laser is equivalent to inverting its propagation direction or inverting
$\vec{B}_{0}$. Therefore, in this letter, we take the $z$ axis as the
quantization axis and focus on the $\sigma ^{+}$-polarized pump with the
pump laser's orientation angle $\theta \in \lbrack 0,\pi /2]$. For the $%
\sigma ^{-}$-polarized pump, we only need to change $B_{0}$ to $-B_{0}$ in
the calculation. For the D1 transition, in the rotating frame with respect
to the pump laser's frequency, the master equation for the alkali-metal atom
is \cite{happer2010optically,PhysRevA.58.1412,PhysRevA.82.043417}:%
\begin{equation}
\partial _{t}\rho =-i\left[ H,\rho \right] +\mathcal{L}_{PP}\rho +\mathcal{L}%
_{SP}\rho +\mathcal{L}_{SS}\rho .  \label{1}
\end{equation}%
Here, the Hamiltonian $H=H_{HF}+H_{B}+H_{LA}+H_{D}$, where $H_{HF}$ is the
hyperfine interaction%
\begin{eqnarray}
H_{HF} &=&\sum_{m}\Delta _{S}\left\vert am\right\rangle _{SS}\left\langle
am\right\vert +\Delta \left\vert am\right\rangle _{PP}\left\langle
am\right\vert  \notag \\
&&+\left( \Delta -\Delta _{P}\right) \left\vert bm\right\rangle
_{PP}\left\langle bm\right\vert
\end{eqnarray}%
with the hyperfine splitting $\Delta _{S}$ ($\Delta _{P}$) in the ground
states $\left\vert Fm\right\rangle _{S}$ (excited states $\left\vert
Fm\right\rangle _{P}$), $F=a$, $b$, and the detuning $\Delta $ of the pump
laser (see Fig.~\ref{fig1}(b)); $H_{B}$ depicts the interaction between the
spins and the external magnetic field%
\begin{equation}
H_{B}=g_{e}\mu _{\mathrm{B}}B_{z}S_{z}+g_{I}\mu _{\mathrm{N}}B_{z}I_{z}
\end{equation}%
with the electron (nuclear) spin operator$\vec{S}$ ($\vec{I}$) and the
electron (nuclear) g-factor $g_{e}$ ($g_{I}$) of the alkali-metal atom; $%
H_{LA}$ is the light-atom interaction%
\begin{equation}
H_{LA}=-\frac{E_{0}}{2}\left( \frac{1}{\sqrt{2}}\sum_{\sigma =\pm
1}d_{\sigma }\left( \cos \theta +\sigma \right) +d_{z}\sin \theta \right)
\label{5}
\end{equation}%
with the electric field $E_{0}$ of the pump laser and the atom's dipole
moment $\vec{d}$ while $d_{\pm 1}\equiv -\left( d_{x}\pm id_{y}\right) /%
\sqrt{2}$. Note that under the rotating-wave approximation, the dipole
moment $d_{\pm }$ has only matrix elements between $\left\vert
Fm\right\rangle _{S}$ and $\left\vert Fm\pm 1\right\rangle _{P}$ \cite%
{Walls2008}. The coupling to the driving field $H_{D}=g_{e}\mu _{\mathrm{B}%
}B_{1}\left( S_{x}\cos \theta -S_{z}\sin \theta \right) \cos \omega t$
induces polarization in the $xy$ plane. In experiments, the precession
frequency is determined by the zero crossing $\omega _{0}$ of the in-phase
part in $\left\langle S_{x}\cos \theta +S_{z}\sin \theta \right\rangle $
when the probe laser is parallel to the RF field ($\omega _{0,\parallel }$)
or out-of-phase part in $\left\langle S_{y}\right\rangle $ when the probe
laser is perpendicular to the RF field ($\omega _{0,\perp }$). Apart from
the coherent dynamics, the alkali-metal atom experiences excited-state
mixture \cite{happer2010optically,PhysRevA.82.043417} ($\mathcal{L}_{PP}$)
and quenching \cite{happer2010optically,PhysRevA.82.043417} ($\mathcal{L}%
_{SP}$)\ caused by collisions between alkali-metal atoms in excited states
and buffer gas atoms, and dissipation \cite%
{happer2010optically,PhysRevA.58.1412} ($\mathcal{L}_{SS}$)$\mathcal{\ }$in
the ground states induced by collisions between alkali-metal atoms.

Within the geophysical range, the interaction $H_{B}$ can be treated as a
perturbation to the hyperfine states. To the second, the energy $E\left(
a/b,m\right) $ (apart from a constant independent of $B_{0}$) of the
ground-state Zeeman sublevel $\left\vert a\text{/}b\text{, }m\right\rangle
_{S}$ is%
\begin{equation}
E\left( a/b,m\right) \approx \left( \pm \mu _{\mathrm{eff}}-g_{I}\mu _{%
\mathrm{N}}\right) B_{0}m\mp \omega _{\mathrm{rev}}m^{2},  \label{2}
\end{equation}%
where the effective magneton $\mu _{\mathrm{eff}}\equiv \left( g_{S}\mu _{%
\mathrm{B}}+g_{I}\mu _{\mathrm{N}}\right) /\left( 2I+1\right) $ and the
quantum-beat revival frequency $\omega _{\mathrm{rev}}\equiv \mu _{\mathrm{%
eff}}^{2}B_{0}^{2}/\Delta _{S}$. The last term in Eq.~(\ref{2}) is the
lowest-order NLZ splitting \cite{PhysRevA.79.023406,PhysRevA.82.023417}.
With this NLZ effect,\ the energy spacing between two adjacent Zeeman
sublevels depends on the magnetic quantum number $m$, as shown in Fig.~\ref%
{fig1}(c), leading to heading errors: the population in each state $%
\left\vert Fm\right\rangle _{S}$ changes when orientation angle $\theta $
varies and thus the measured precession frequency $\omega _{0}$ is dependent
on $\theta $. Another contribution to heading errors is the LS \cite%
{PhysRevA.99.013420} that shifts the energy of the state $\left\vert
Fm\right\rangle _{S}$ by an amount depending on $m$ and $\theta $. Apart
from the NLZ effect and LS, we find that the NuZ effect also causes heading
errors. Fig.~\ref{fig1}(c) shows that for the same $m$, the adjacent-states'
energy level spacings in the $a$ and $b$ manifolds are different because of
the linear NuZ splitting $\omega _{\mathrm{NuZ}}\equiv g_{I}\mu _{\mathrm{N}%
}\left\vert B_{0}\right\vert $. As $\theta $ varies, the populations in the $%
a$ and $b$ manifolds change, shifting the precession frequency. Note that
inverting the helicity of the pump is equivalent to inverting $\vec{B}_{0}$,
so we plot the energy spacings in Fig.~\ref{fig1}(c) for $\pm B_{0}$ to
illustrate physical insights for the opposite-helicity case.

\begin{figure}[tb]
\begin{center}
\includegraphics[width=\linewidth]{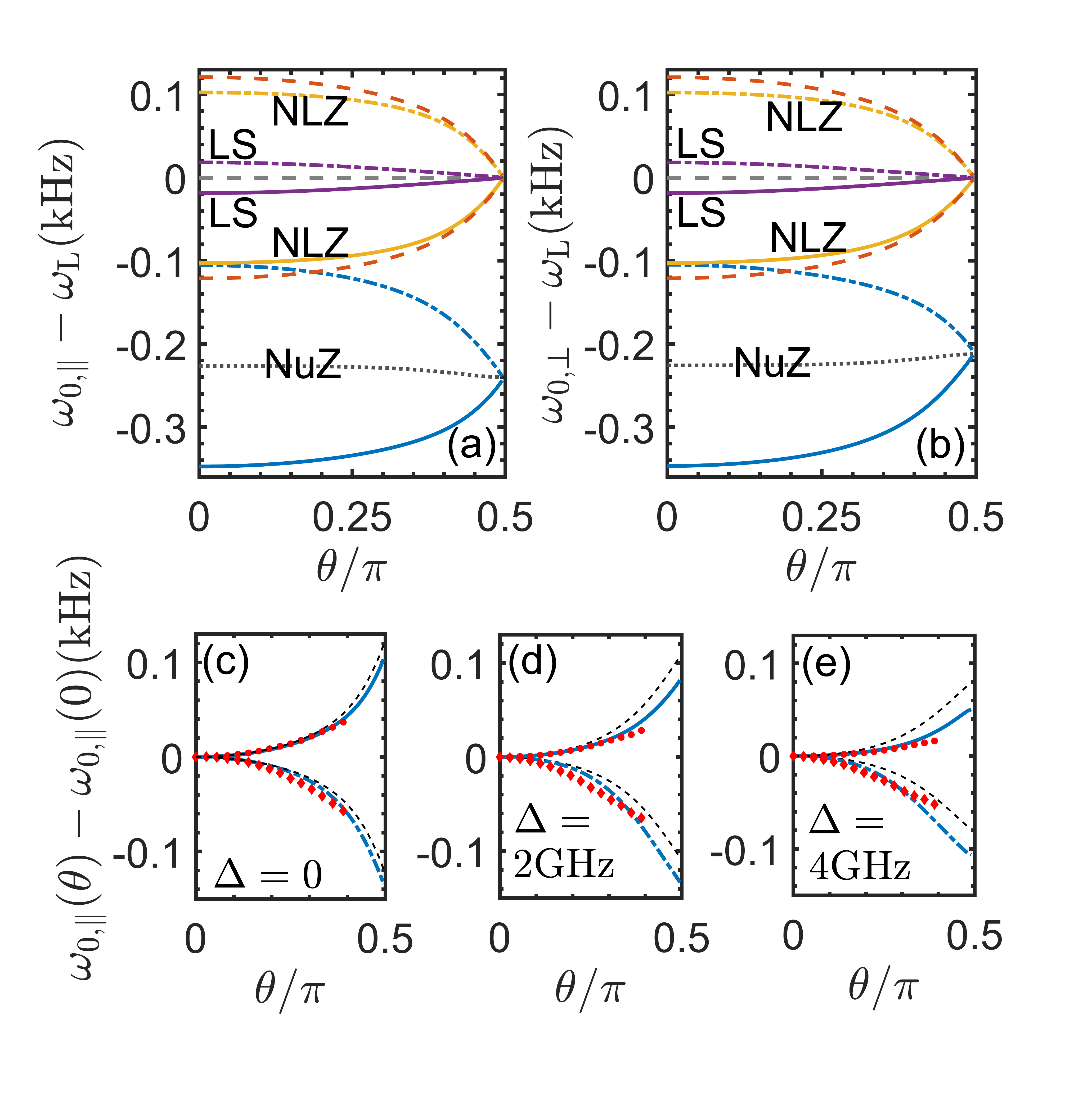}
\end{center}
\caption{(a) Contributions to the precession frequency $\protect\omega _{0}$
(shown as the deviation from the Larmor frequency $\protect\omega _{\mathrm{L%
}}$) from the NLZ effect, LS, and the NuZ effect, with the probe laser
parallel (a) or perpendicular (b) to the RF field. The detuning $\Delta $ of
the pump laser is $0$. Here and after, if not specified, solid lines are for
$\protect\sigma ^{+}$ polarization while dotted-dashed lines are for the
negative $\protect\sigma ^{-}$ case. The precession frequencies when
including all the three contributions are shown at the bottom as the blue
lines. Their average value $\left[ \protect\omega _{0}^{+}\left( \protect%
\theta \right) +\protect\omega _{0}^{-}\left( \protect\theta \right) \right]
/2-\protect\omega _{\mathrm{L}}$ is just the precession frequency when
including only the NuZ effect (black dotted line). For comparison, the
precession frequencies when neglecting the NuZ effect and its average values
are shown as the red and flat grey dash lines respectively in the upper
part. (c)-(e) Heading errors $\protect\omega _{0,\parallel }\left( \protect%
\theta \right) -\protect\omega _{0,\parallel }\left( 0\right) $ for
different detuning $\Delta $. The results without the NuZ effect are shown
as the grey dash lines, while the experimental data are plotted in dots ($%
\protect\sigma ^{+}$) and diamonds ($\protect\sigma ^{-}$). }
\label{fig2}
\end{figure}

\begin{figure}[tb]
\begin{center}
\includegraphics[width=\linewidth]{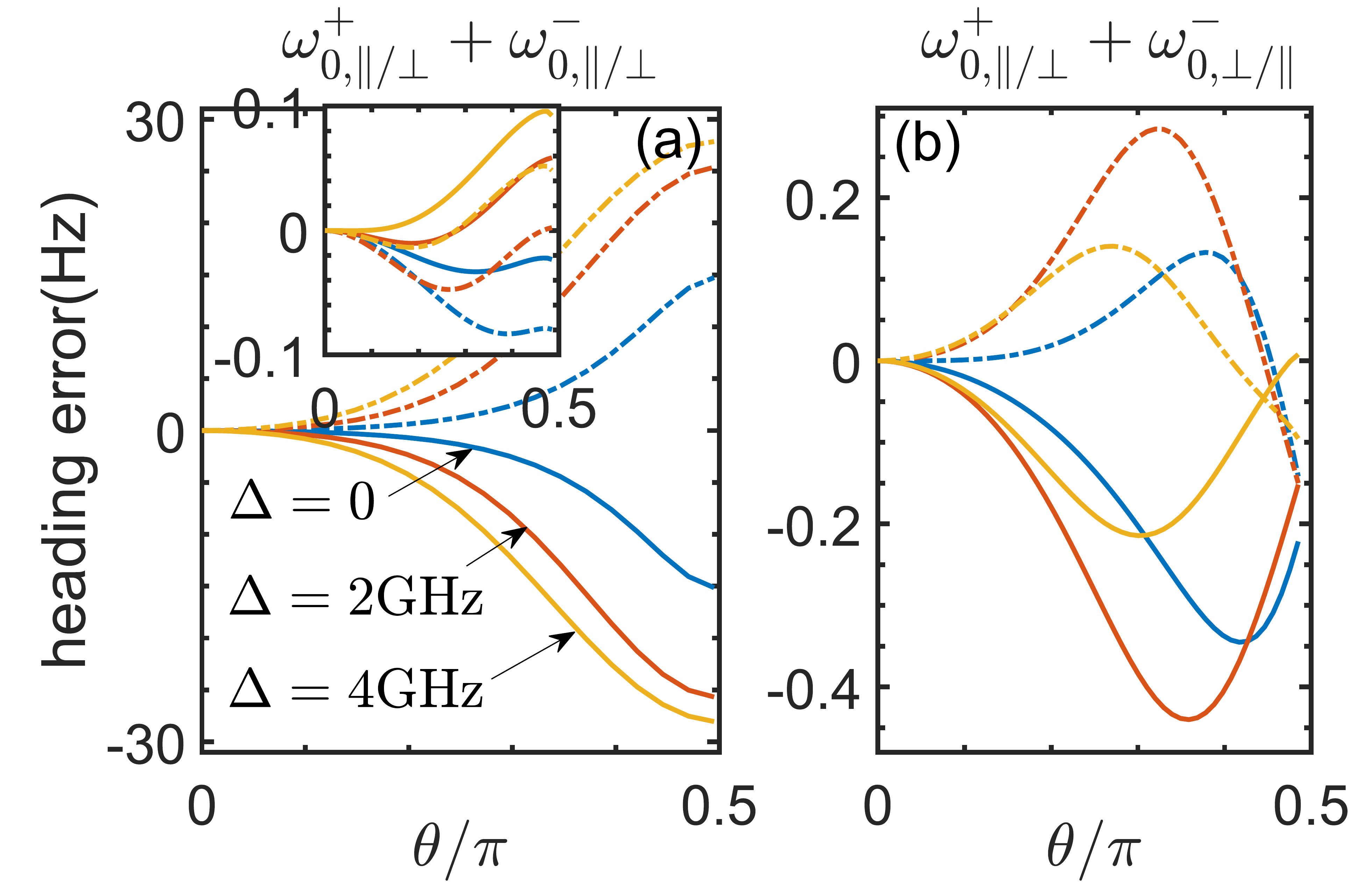}
\end{center}
\caption{Heading errors in the average of two magnetometers pumped by lasers
with opposite helicities. Parameters are the same as in Fig.~\protect\ref%
{fig2}. (a) Solid lines are for $\protect\omega _{0,\parallel }^{+}+\protect%
\omega _{0,\parallel }^{-}$ while dotted-dash lines are for $\protect\omega %
_{0,\perp }^{+}+\protect\omega _{0,\perp }^{-}$. The inset shows the result
without the NuZ effect. (b) Solid lines are $\protect\omega _{0,\parallel
}^{+}+\protect\omega _{0,\perp }^{-}$ while dotted dashed lines are for $%
\protect\omega _{0,\perp }^{+}+\protect\omega _{0,\parallel }^{-}$.}
\label{fig3}
\end{figure}

With the experimental condition: in a magnetic shield, a $4^{3}$mm$^{3}\
\left. ^{85}\text{Rb}\right. $ cell with $700$Torr N$_{2}$ is heated to $90$
Celsius, the pump laser's power is around $50$~$\mathrm{\mu W}$, and the
probe laser is about $0.2$nm detuned from the D1 transition, we solve
numerically the master equation (\ref{1}) by adiabatically eliminating the
excited states \cite{Gardiner2004} and applying the linear response theory
\cite{fetter2012quantum} for the RF driving \cite{PhysRevA.99.063411}. The
deviation of the precession frequency $\omega _{0,\parallel /\perp }\left(
\theta \right) $ from the Larmor frequency $\omega _{\mathrm{L}}$ is plotted
in Fig.~\ref{fig2}(a) and (b) for $B_{0}=55\mathrm{\mu T}$, with the pump
laser's detuning $\Delta =0$. It shows that: (1) the precession frequency is
smaller than $\omega _{\mathrm{L}}$, since in the $a$ manifold that has more
population than the $b$ manifold because of the optical pumping, the linear
NuZ effect contributes $-\omega _{\mathrm{NuZ}}$ to the precession frequency
and this contribution is larger than the NLZ effect and LS; (2) the
precession frequency $\omega _{0,\parallel /\perp }^{+}\left( \theta \right)
$ for the $\sigma ^{+}$ polarization is smaller than $\omega _{0,\parallel
/\perp }^{-}\left( \theta \right) $ for the $\sigma ^{-}$ case\ because of
the sign of the NLZ splitting for the $m>0$ states which has more
populations than the $m<0$ ones; (3) as $\theta $ increases, $\omega
_{0,\parallel /\perp }^{+}\left( \theta \right) $ increases ($\omega
_{0,\parallel /\perp }^{-}\left( \theta \right) $ decreases) because states
with smaller $m$ and states in the $b$ manifolds get more populated and
these states have lager (smaller) precession frequencies; (4) the heading
errors for opposite helicity pumps are not symmetric \cite%
{doi:10.1063/1.1663412}, i.e., for two given angles $\theta _{1}$ and $%
\theta _{2}$,%
\begin{equation}
\omega _{0}^{+}\left( \theta _{1}\right) -\omega _{0}^{+}\left( \theta
_{2}\right) \neq \omega _{0}^{-}\left( \theta _{2}\right) -\omega
_{0}^{-}\left( \theta _{1}\right) ,  \label{3}
\end{equation}%
which holds for both the parallel and vertical probe lasers. We note that
this asymmetry was also presented \cite{doi:10.1063/1.1663412}. In the
following, we will show that this asymmetry is resulted from the NuZ effect.

The precession frequencies from the three sources: the NLZ effect, the LS,
and the NuZ are also separately shown. We see that both the NLZ effect and
the LS lead to symmetric heading errors for positive and negative $B_{0}$,
and they average to the Larmor frequency \cite{symmetry,supmat}. However,
the precession frequency results in the same precession frequencies for both
the $\sigma ^{\pm }$-polarization cases \cite{supmat}, which is just the
angular dependence of the average value $\left( \omega _{0}^{+}+\omega
_{0}^{-}\right) /2-\omega _{\mathrm{L}}$ when including all three effects,
and results in smaller heading error for $\sigma ^{+}$ polarization with
parallel probe lasers but larger with vertical probe lasers.

The total heading error $\omega _{0,\parallel }\left( \theta \right) -\omega
_{0,,\parallel }\left( 0\right) $ with the parallel probe laser is shown in
Fig.~\ref{fig2}(c)-(e) for different detunings $\Delta =0$, $2\mathrm{GHz}$,
and $4\mathrm{GHz}$, respectively. For comparison, the results without the
NuZ effect are also plotted. When $\Delta =0$, the $b$-manifold is
resonantly pumped and its population is small. Therefore, the NuZ effect is
not significant. However, when $\Delta $ increases and more populations are
in the $b$-manifold, the heading errors with and without the NuZ effect
deviate a lot, and the experimental data agrees with the former one \cite%
{experiment}.

Considering the heading-error asymmetry that is shown to be induced by the
NuZ effect, one can choose the orientation of the sensor, the polarization
of the pump laser, or the direction of the probe laser with respect to the
driving field to take advantage of the smaller angular-dependent case.
Nonetheless, the heading error is yet too large for some applications such
as an aeromagnetic survey. There have been many proposals in the literature
to suppress the heading error \cite%
{PhysRevA.75.051407,PhysRevLett.120.033202,PhysRevA.82.023417,scholtes2012light,PhysRevA.79.023406,lee2021heading}%
, but none of them could compensate the contribution from the NuZ effect (in
a recent work \cite{lee2021heading}, a correction method is proposed to
reduce the heading errors where the NuZ effect is considered. In contrast to
ours, their system is pumped by short pulses and nonlinear but not the
linear NuZ effect contributes to its orientation-dependent precession
frequency). Thus, the suppression of the heading error becomes less
efficient at larger tilted angles where the atomic population in the $b$
manifold becomes more prominent. For instance, one of the common methods is
to average the output precession frequencies from cells pumped by lasers
with opposite helicities \cite%
{doi:10.1063/1.1663412,scholtes2012light,PhysRevA.99.013420} and probe light
in the same directions to the driving field. However, because of the
asymmetry, the average value in this scheme still has a considerable angular
dependence that can not be neglected in precise measurements. As shown in
Fig.~\ref{fig3}(a), the heading errors in a full range of the orientation
angles $\theta $ are $15\sim 28$Hz depending on the detuning $\Delta $ (in
contrast, the heading error without the NuZ effect, shown in the inset, is
negligible). With lager $\Delta $ or larger $\theta $, the population in the
$b$-manifold increases and the NuZ effect induces bigger heading errors. To
compensate for the NuZ-effect-induced heading error in this two-pump scheme,
we propose to employ one probe laser parallel to the driving field and
another one perpendicular to it \cite{supmat}. As shown in Fig.~\ref{fig3}%
(b), the heading errors for different detunings $\Delta $ are within $0.5$%
~Hz.

\begin{figure}[tb]
\begin{center}
\includegraphics[width=\linewidth]{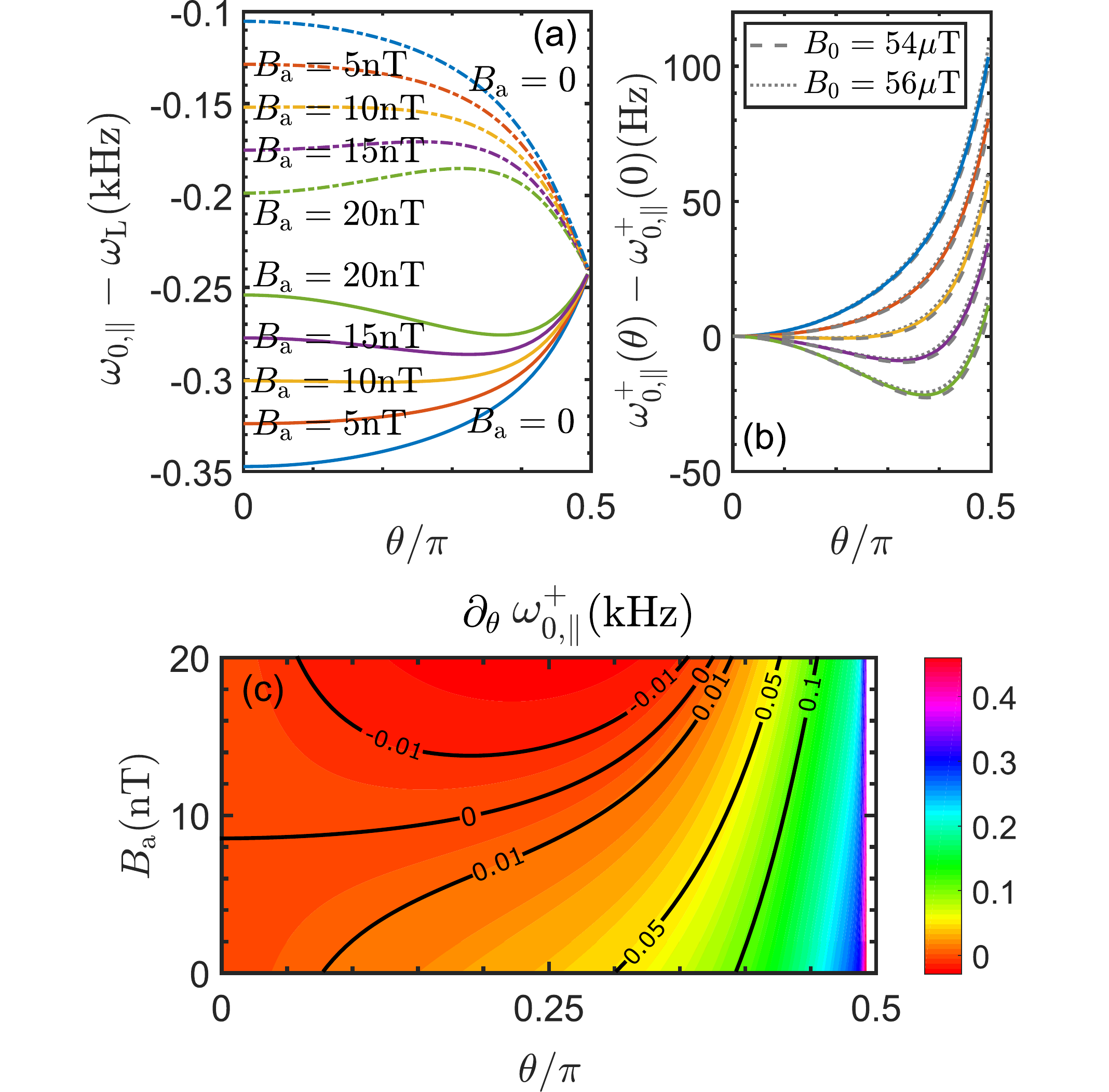}
\end{center}
\caption{(a) The relative precession frequency $\protect\omega _{0,\parallel
}-\protect\omega _{\mathrm{L}}$ with auxiliary field $B_{\mathrm{a}}=0$, $10$%
, $20$nT for $\protect\sigma ^{\pm }$ pumps. (b) The heading error $\protect%
\omega _{0,\parallel }^{+}\left( \protect\theta \right) -\protect\omega %
_{0,\parallel }^{+}\left( 0\right) $ and (c) the derivative $\partial _{%
\protect\theta }\protect\omega _{0,\parallel }^{+}$. }
\label{fig4}
\end{figure}

The method using opposite-helicity lasers requires an exact match of
parameters in the two magnetometers, which may cause difficulties in
practice. Therefore, we propose another way to suppress the heading error by
utilizing an auxiliary magnetic field $\vec{B}_{\mathrm{a}}$ in parallel
(for $\sigma ^{+}$ pump) or antiparallel (for $\sigma ^{-}$ pump) to the
propagation direction of the pump laser. Its magnitude $B_{\mathrm{a}}$ is a
small positive constant (much smaller than $B_{0}$) so that the NLZ and the
NuZ effects induced by $B_{a}$ can be neglected. Without loss of generality,
we consider the parallel probe laser case. With the auxiliary field $B_{%
\mathrm{a}}=0$, $10$, $20\mathrm{nT}$, the relative precession frequency $%
\omega _{0}-\omega _{\mathrm{L}}$ is plotted in Fig.~\ref{fig4}(a), while
other parameters are the same as in Fig.~\ref{fig2}(a). In presence of the
auxiliary field $B_{\mathrm{a}}$, $\omega _{0}$ has a positive/negative
offset for the $\sigma ^{+}$/$\sigma ^{-}$ polarization since the connection
between $\omega _{0}\left( \theta ,B_{\mathrm{a}}=0\right) $ and $\omega
_{0}\left( \theta ,B_{\mathrm{a}}\neq 0\right) $ can be approximated as%
\begin{equation}
\omega _{0}\left( \theta ,B_{\mathrm{a}}\right) \approx \omega _{0}\left(
\theta ,B_{\mathrm{a}}=0\right) +\mathrm{sign}\left( \sigma ^{\pm }\right)
\mu _{\mathrm{eff}}B_{\mathrm{a}}\cos \theta ,
\end{equation}%
where $\mathrm{sign}\left( \sigma ^{\pm }\right) =\pm 1$ corresponding to
the $\sigma ^{\pm }$-polarization. The heading error in each case becomes
smaller, i.e., to achieve the same accuracy, the required accuracy of the
angle $\theta$ is relaxed, because the angular dependence of $\omega
_{0}\left( \theta ,B_{\mathrm{a}}=0\right) $ can be approximated by $\mp
\cos \theta $ for $\sigma ^{\pm }$-polarization \cite{supmat}. Taking the $%
\sigma ^{+}$ pump as an example, we plot\ the heading error $\omega
_{0}\left( \theta \right) -\omega _{0}\left( \theta =0\right) $ in Fig.~\ref%
{fig3}(b) for $B_{0}=54\mathrm{\mu T}$, $55\mathrm{\mu T}$, and $56\mathrm{%
\mu T}$ (different magnetic field strengths are considered since in
practice, the strength of the circumstance is roughly known). It shows that
the heading error with a change of $2\%$ of the external field is nearly
unchanged. Thus, our method does not require a precise measurement of $B_{0}$%
. Furthermore, there exists an angle $\theta _{0}$ at which the derivative $%
\partial _{\theta }\omega _{0}\left( \theta _{0}\right) =0$. As shown in
Fig.~\ref{fig4}(c), $\theta _{0}=0$ when $B_{\mathrm{a}}$ is smaller than $%
8.2$nT,$\ $and after that, $\theta _{0}$ is a monotonic function of $B_{%
\mathrm{a}}$. Therefore, the magnitude of the auxiliary field $B_{\mathrm{a}%
} $ can be tuned to flatten the heading error curve around a demanded angle $%
\theta _{0}$. In practical use, $B_{\mathrm{a}}$ can be determined after the
heading-error curve is obtained.

We have presented a full analysis of the NuZ effect on the heading errors in
atomic magnetometers. Our theoretical result acquired from numerically
solving the master equation agrees with our experimental data. Based on our
study, one can design the magnetometer to have it work in the smaller
heading error regime. Considering the NuZ effect, we suggest to suppress the
heading error by employing pump lasers with opposite helicities and probe
lasers propagating in different directions (one parallel and another
perpendicular to the driving fields). Furthermore, we propose a scheme to
reduce the heading errors by utilizing a small magnetic field
parallel/antiparallel to the propagation direction of the pump laser. The
magnitude of this auxiliary field can be tuned to flatten the heading error
around the desired angle, which has promising applications for magnetometers
working around a certain orientation.

The authors acknowledge support by the National Natural Science Foundation
of China Grants No.61627806 and No.61903045.

\bibliographystyle{apsrev}
\bibliography{ref}

\begin{thebibliography}{33}
\bibitem[{sym()}]{symmetry}
\bibinfo{note}{In this equality, we have neglected the corrections in the
  Hyperfined states from the external magnetic field. When including it,
  without the NuZ effet, the average value slightly deviates from the Larmor
  frequency, as shown in the inset in Fig. 3(a)}.
\bibitem[{exp()}]{experiment}
\bibinfo{note}{In our experiment, instead of changing the helicity of the pump
  laser, we invert the external field, which can be proved to be equivalent.
  Compared with the good agreement between the theoretical and experimental
  results in the left-handed-circular-polarization case, the relatively larger
  deviation in the right-handed case is probably because of the change of
  system parameters in the experiment when the external field is inverted}.
\end{thebibliography}
%\bibliography{v5.bbl}

\newpage \widetext

\begin{center}
\textbf{\large Supplemental Material: \\[0pt]
Nuclear Zeeman effect on heading errors and the suppression in atomic
magnetometers}
\end{center}

\setcounter{equation}{0} \setcounter{figure}{0} %\setcounter{table}{0}
\makeatletter

\renewcommand{\thefigure}{SM\arabic{figure}} \renewcommand{\thesection}{SM%
\arabic{section}} \renewcommand{\theequation}{SM\arabic{equation}}

In this supplemental material, we provide detailed proof of the system's
symmetry, show the nuclear Zeeman (NuZ) effect on the heading errors under
various sets of system parameters including explaining analytically the
asymmetric heading errors induced by the NuZ effect, and compare the two
schemes for suppressing the heading errors by using two atomic cells pumped
by opposite-helicity lasers and probed by lasers propagating in the same or
orthogonal directions (parallel or perpendicular to the RF driving field).
At last, we give a physical insight into our auxiliary-field method and
compare it with other approaches in the literature.

\section{Symmetries in the system}

\subsection{Master equation}

The master equation for the alkali-metal atoms is

\begin{equation}
\partial _{t}\rho =-i\left[ H,\rho \right] +\mathcal{L}_{PP}\rho +\mathcal{L}%
_{SP}\rho +\mathcal{L}_{SS}\rho ,
\end{equation}%
where the Hamiltonian $H=H_{HF}+H_{B}+H_{LA}+H_{D}$. The light-atom
interaction for $\sigma ^{+}$-polarized lasers \cite{Walls2008}%
\begin{eqnarray}
H_{LA} &=&-\frac{E_{0}}{2}\left( \frac{1}{\sqrt{2}}\sum_{\sigma =\pm
1}d_{\sigma }\left( \cos \theta +\sigma \right) +d_{z}\sin \theta \right)
\notag \\
&=&-\Omega \left[ \left( \cos \theta +1\right) A_{+}+\left( \cos \theta
-1\right) A_{-}+\left( A_{0+}+A_{0-}\right) \sin \theta \right] +H.c.,
\end{eqnarray}%
where the Rabi frequency%
\begin{equation}
\Omega =\frac{E_{0}}{2\sqrt{3}}\left\langle 1/2\right\Vert d\left\Vert
1/2\right\rangle ,
\end{equation}%
$A_{0\pm }=\left\vert \pm \right\rangle _{SP}\left\langle \pm \right\vert $,
and $A_{\pm }=\left\vert \mp \right\rangle _{SP}\left\langle \pm \right\vert
$, with the notations of the fine states $\left\vert \pm \right\rangle
_{S\left( P\right) }\equiv \left\vert ^{2}S_{1/2}\left( ^{2}P_{1/2}\right)
,\pm \frac{1}{2}\right\rangle $. The dissipation from the excited-state
mixture resulting from collisions between alkali-metal atoms and buffer gas
is depicted by \cite{happer2010optically,PhysRevA.82.043417}%
\begin{equation}
\mathcal{L}_{PP}\rho =\gamma _{Mix}\left( 2\vec{J}^{P}\cdot \rho \vec{J}%
^{P}-\left\{ \rho ,\vec{J}^{P}\cdot \vec{J}^{P}\right\} \right) ,
\end{equation}%
where $\gamma _{Mix}$ is the collision rate of excited atoms with the buffer
gas, $\boldsymbol{J}^{P}$ is the angular momentum of the excited atoms
defined as $J_{x}^{P}=\left( \left\vert +\right\rangle _{PP}\left\langle
-\right\vert +\left\vert -\right\rangle _{PP}\left\langle +\right\vert
\right) /2$, $J_{y}=\left( \left\vert +\right\rangle _{PP}\left\langle
-\right\vert -\left\vert -\right\rangle _{PP}\left\langle +\right\vert
\right) /2i$, and $J_{z}=\left( \left\vert +\right\rangle _{PP}\left\langle
+\right\vert -\left\vert -\right\rangle _{PP}\left\langle -\right\vert
\right) /2$. The buffer gas also induces decay to the ground states \cite%
{happer2010optically,PhysRevA.82.043417}%
\begin{equation}
\mathcal{L}_{SP}\rho =\Gamma _{Q}\left( \sum_{j=0\pm ,\pm }2A_{j}\rho
A_{j}^{\dagger }-\left\{ \rho ,A_{j}^{\dagger }A_{j}\right\} \right) ,
\end{equation}%
where $\Gamma _{Q}$ is the quenching rate. Here, the spontaneous decay is
neglected since its rate is much smaller than $\Gamma _{Q}$ under our
experimental condition (several hundreds Torr N$_{2}$). Dissipation $%
\mathcal{L}_{SS}\rho $ in the ground-state is \cite%
{happer2010optically,PhysRevA.58.1412}%
\begin{eqnarray}
\mathcal{L}_{SS}\rho &=&\left( \gamma _{SD}+\gamma _{SE}\right) \left( 2\vec{%
S}\cdot \rho \vec{S}\boldsymbol{-}\left\{ \rho ,\vec{S}\cdot \vec{S}\right\}
\right) +2\gamma _{SE}\left\langle S_{z}\right\rangle \left( S_{+}\rho
S_{-}-S_{-}\rho S_{+}\boldsymbol{+}\left\{ \rho ,S_{z}\right\} \right)
\notag \\
&&+2\gamma _{SE}\left\langle S_{+}\right\rangle \left( S_{-}\rho
S_{z}-S_{z}\rho S_{-}\boldsymbol{+}\frac{1}{2}\left\{ \rho ,S_{-}\right\}
\right) +H.c.,
\end{eqnarray}%
where $\gamma _{SD}$ ($\gamma _{SE}$) is the spin-destruction (exchange)
rate.

\subsection{Linear response}

The master equation (\ref{1}) can be written as $\partial _{t}\rho =\left(
\mathcal{L}_{0}+\mathcal{L}_{1}\right) \rho $, where%
\begin{equation}
\mathcal{L}_{0}\rho =-i\left[ H_{HF}+H_{B}+H_{LA},\rho \right] +\mathcal{L}%
_{PP}\rho +\mathcal{L}_{SP}\rho +\mathcal{L}_{SS}\rho
\end{equation}%
and%
\begin{equation}
\mathcal{L}_{1}\rho =-i\left[ H_{D},\rho \right] .
\end{equation}%
Since the RF field is weak, one can treat $\mathcal{L}_{1}$ as a
perturbation and keep it to the first order \cite{fetter2012quantum}. As a
result, the density matrix
\begin{equation}
\rho =\rho _{0}+\rho _{1}^{\left( +\right) }e^{i\omega t}+\rho _{1}^{\left(
-\right) }e^{-i\omega t},
\end{equation}%
where in the long-term limit, $\mathcal{L}_{0}\rho _{0}=0$,
\begin{equation}
\left( \mathcal{L}_{0}\mp i\omega \right) \rho _{1}^{\left( \pm \right) }+%
\mathcal{L}_{1}\rho _{0}=0,  \label{sm3}
\end{equation}%
and $\rho _{1}^{\left( -\right) }=\rho _{1}^{\left( +\right) \dag }$. Here,
we have removed the time dependence in $\mathcal{L}_{1}$ and redefined it as%
\begin{equation}
\mathcal{L}_{1}\rho =-ig_{e}\mu _{\mathrm{B}}B_{1}\left[ \left( S_{x}\cos
\theta -S_{z}\sin \theta \right) ,\rho \right] .
\end{equation}%
Note that in $\mathcal{L}_{0}\rho _{1}^{\left( +\right) }$, the collission
induced dissipation $\mathcal{L}_{SS}\rho _{1}^{\left( +\right) }$ is%
\begin{eqnarray}
\mathcal{L}_{SS}\rho _{1}^{\left( +\right) } &=&\left( \gamma _{SD}+\gamma
_{SE}\right) \left( 2\vec{S}\cdot \rho _{1}^{\left( +\right) }\vec{S}%
\boldsymbol{-}\left\{ \rho _{1}^{\left( +\right) },\vec{S}\cdot \vec{S}%
\right\} \right) +2\gamma _{SE}\mathrm{Tr}\left( S_{z}\rho _{0}\right)
\left( S_{+}\rho _{1}^{\left( +\right) }S_{-}-S_{-}\rho _{1}^{\left(
+\right) }S_{+}\boldsymbol{+}\left\{ \rho _{1}^{\left( +\right)
},S_{z}\right\} \right)  \notag \\
&&+2\gamma _{SE}\left\langle S_{+}\rho _{1}^{\left( +\right) }\right\rangle
\left( S_{-}\rho _{0}S_{z}-S_{z}\rho _{0}S_{-}\boldsymbol{+}\frac{1}{2}%
\left\{ \rho _{0},S_{-}\right\} \right)  \notag \\
&&+2\gamma _{SE}\left\langle S_{-}\rho _{1}^{\left( +\right) }\right\rangle
\left( S_{z}\rho _{0}S_{+}-S_{+}\rho _{0}S_{z}\boldsymbol{+}\frac{1}{2}%
\left\{ \rho _{0},S_{+}\right\} \right) .
\end{eqnarray}%
When the probe laser is parallel to the RF field, the precession frequency $%
\omega _{0}$ is determined by the zero-crossing of the in-phase frequency
response
\begin{equation}
S_{\parallel }\left( \omega \right) \equiv 2\mathrm{Re}\left[ \mathrm{Tr}%
\left( \left( S_{x}\cos \theta -S_{z}\sin \theta \right) \rho _{1}^{\left(
+\right) }\right) \right] =2\mathrm{Re}\left[ \mathrm{Tr}\left( \left(
S_{x}\cos \theta -S_{z}\sin \theta \right) \rho _{1}^{\left( -\right)
}\right) \right] ,
\end{equation}%
while when the probe laser is perpendicular to the RF field, $\omega _{0}$
is determined by the zero-crossing of the out-of-phase output
\begin{equation}
S_{\perp }\left( \omega \right) \equiv 2\mathrm{Im}\left[ \mathrm{Tr}\left(
S_{y}\rho _{1}^{\left( +\right) }\right) \right] =-2\mathrm{Im}\left[
\mathrm{Tr}\left( S_{y}\rho _{1}^{\left( -\right) }\right) \right] .
\end{equation}

To prove $\omega _{0}$ is an even function of the tilted angle $\theta $, we
perform a rotation of angle $2\theta $ along the $y$ axis to the atomic
system. Under this unitary transformation, the light-atom interaction $%
H_{LA}\left( \theta \right) $ becomes $H_{LA}\left( -\theta \right) $, the
driving term $H_{D}\left( \theta \right) $ becomes $-H_{D}\left( -\theta
\right) $, while other parts in the master equation are invariant. As a
result, the first-order density matrix $\rho _{1}^{\left( +\right) }=-\left(
\mathcal{L}_{0}-i\omega \right) ^{-1}\mathcal{L}_{1}\rho _{0}$ with positive
frequency changes to $-\rho _{1}^{\left( +\right) }\left( -\theta \right) $,
and thus$\ S_{\parallel /\perp }\left( \theta ,\omega \right) =S_{\parallel
/\perp }\left( -\theta ,\omega \right) $. Therefore, the precession
frequency $\omega _{0}$ is invariant when changing $\theta $ to $-\theta $.

Similarly, performing a rotation of angle $\pi $ along the $y$ axis and a $%
Z_{2}$ transformation to the excited states so that each excited state has
an additional global phase $\pi $, the interaction to the external magnetic
field $H_{B}\left( B_{0}\right) $ changes to $H_{B}\left( -B_{0}\right) $,
and the light-matter interaction in Eq.~(4) becomes%
\begin{equation}
H_{LA}=-\frac{E_{0}}{2}\left( \frac{1}{\sqrt{2}}\sum_{\sigma =\pm
1}d_{-\sigma }\left( \cos \theta +\sigma \right) +d_{z}\sin \theta \right) ,
\end{equation}%
i.e., the polarization of the pump light changes. The driving $H_{D}$
changes to $-H_{D}$. Therefore, the precession frequency $\omega _{0}$\ that
is determined by the zero crossings of the in-phase/out-of-phase part $%
S_{\parallel /\perp }$ is invariant when changing the pump laser's
polarization and at the same time inverting the magnetic field $B_{0}$, or
equivalently,\ inverting the helicity of the pump laser is equivalent to
inverting the magnetic field $B_{0}$.

\section{Nuclear Zeeman effect on heading errors}

\subsection{Adiabatic elimination of the excited states}

In the steady state $\rho _{0}$, populations in the excited states are
negligible because their decay rates is much larger than the Rabi frequency.
Therefore, in the zero-order master equation $\partial _{t}\rho =\mathcal{L}%
_{0}\rho $,\ we can adiabatic eliminate the excited state and acquire an
effective master equation in the ground-state subspace \cite%
{Gardiner2004,PhysRevA.99.063411}. For this purpose, we rewrite the Lindblad
operator $\mathcal{L}_{0}$ as $\mathcal{L}_{0}=\mathcal{L}_{0}^{\left(
0\right) }+\mathcal{L}_{0}^{\left( 1\right) }$, where%
\begin{equation}
\mathcal{L}_{0}^{\left( 0\right) }\rho =-i\left[ H_{hf}+H_{B},\rho \right] +%
\mathcal{L}_{PP}\rho +\mathcal{L}_{SP}\rho +\mathcal{L}_{SS}\rho
\end{equation}%
and%
\begin{equation}
\mathcal{L}_{0}^{\left( 1\right) }\rho =-i\left[ H_{LA},\rho \right] ,
\end{equation}%
and define two projection operators%
\begin{equation}
\mathcal{P}\rho =\sum_{F_{1}m_{1}F_{2}m_{2}}\left\vert
F_{1}m_{1}\right\rangle _{SS}\left\langle F_{1}m_{1}\right\vert \rho
\left\vert F_{2}m_{2}\right\rangle _{SS}\left\langle F_{2}m_{2}\right\vert
\end{equation}%
and $\mathcal{Q}=1-\mathcal{P}$. Consequently, we have%
\begin{equation}
\partial _{t}\mathcal{P}\rho =\mathcal{P\mathcal{L}}_{0}^{\left( 0\right) }%
\mathcal{P}\rho +\mathcal{P\mathcal{L}}_{0}^{\left( 0\right) }\mathcal{Q}%
\rho +\mathcal{PL}_{0}^{\left( 1\right) }\mathcal{Q}\rho
\end{equation}%
and%
\begin{equation}
\partial _{t}\mathcal{Q}\rho =\mathcal{Q}\text{~}\mathcal{\mathcal{L}}%
_{0}^{\left( 0\right) }\mathcal{Q}\rho +\mathcal{Q}\text{~}\mathcal{L}%
_{0}^{\left( 1\right) }\mathcal{P}\rho +\mathcal{Q}\text{~}\mathcal{L}%
_{0}^{\left( 1\right) }\mathcal{Q}\rho .
\end{equation}%
The solution of $\mathcal{Q}\rho $ can be formally written as%
\begin{equation}
\mathcal{Q}\rho \left( t\right) =\int_{0}^{t}e^{\mathcal{Q}\left( \mathcal{%
\mathcal{L}}_{0}^{\left( 0\right) }+\mathcal{L}_{0}^{\left( 1\right)
}\right) \left( t-t^{\prime }\right) }\mathcal{Q}\text{~}\mathcal{L}%
_{0}^{\left( 1\right) }\mathcal{P}\rho \left( t^{\prime }\right) dt^{\prime
},
\end{equation}%
and thus to the second order, the motion equation for $\rho ^{\left(
g\right) }\equiv \mathcal{P}\rho $ is%
\begin{eqnarray}
\partial _{t}\rho ^{\left( g\right) }\left( t\right) &\approx &\mathcal{P%
\mathcal{L}}_{0}^{\left( 0\right) }\rho ^{\left( g\right) }\left( t\right) +%
\mathcal{PL}_{0}^{\left( 1\right) }\int_{0}^{t}e^{\mathcal{Q}\text{ }%
\mathcal{\mathcal{L}}_{0}^{\left( 0\right) }\left( t-t^{\prime }\right) }%
\mathcal{Q}~\mathcal{L}_{0}^{\left( 1\right) }\rho ^{\left( g\right) }\left(
t^{\prime }\right) dt^{\prime }  \notag \\
&&+\mathcal{P\mathcal{L}}_{0}^{\left( 0\right) }\int_{0}^{t}e^{\mathcal{Q}%
\text{ }\mathcal{\mathcal{L}}_{0}^{\left( 0\right) }\left( t-t^{\prime
}\right) }\left( 1+\int_{0}^{t-t^{\prime }}dt^{\prime \prime }e^{-\mathcal{Q}%
\text{ }\mathcal{\mathcal{L}}_{0}^{\left( 0\right) }t^{\prime \prime }}%
\mathcal{Q}\text{~}\mathcal{L}_{0}^{\left( 1\right) }e^{\mathcal{Q}\text{ }%
\mathcal{\mathcal{L}}_{0}^{\left( 0\right) }t^{\prime \prime }}\right)
\mathcal{Q}~\mathcal{L}_{0}^{\left( 1\right) }\rho ^{\left( g\right) }\left(
t^{\prime }\right) dt^{\prime }  \notag \\
&=&\mathcal{L}_{0}^{eff}\rho ^{\left( g\right) }\left( t\right) ,
\label{sm1}
\end{eqnarray}%
where the effective Lindblad operator%
\begin{equation}
\mathcal{L}_{0}^{eff}\rho ^{\left( g\right) }\equiv \mathcal{PL}_{0}\rho
^{\left( g\right) }+\mathcal{PL}_{0}\frac{1}{\mathcal{Q}\text{ }\mathcal{L}%
_{0}}\mathcal{Q}\text{~}\mathcal{L}_{1}\frac{1}{\mathcal{Q}\text{ }\mathcal{L%
}_{0}}\mathcal{Q}~\mathcal{L}_{1}\rho ^{\left( g\right) }-\mathcal{PL}_{1}%
\frac{1}{\mathcal{Q}\text{ }\mathcal{L}_{0}}\mathcal{Q}~\mathcal{L}_{1}\rho
^{\left( g\right) }.  \label{sm2}
\end{equation}%
The imaginary part in the last two terms in Eq. (\ref{sm2}) of $\mathcal{L}%
_{0}^{eff}$ gives the LS.

The Lindblad operator $\mathcal{L}_{0}^{eff}$ shown in Eq.~(\ref{sm2}) can
be obtained numerically in the superspace. The dimension of the effective
master equation in the superspace is $\left( 4I+2\right) ^{2}$. In
principle, one needs to solve $\left( 4I+2\right) ^{2}$ nonlinear equations
since the Zeeman sublevels are mixed due to collisions in the ground states
and pumping-induced dissipations as shown in the last two terms in Eq.~(\ref%
{sm2}). In the geophysical field range, these mixing rates are much smaller
than the Larmor frequency $\omega _{\mathrm{L}}$ under the usual
experimental condition. Therefore, one can ignore the off-diagonal terms in
the steady state $\rho _{0}^{\left( g\right) }$ and consider only $4I+2$
equations. This largely accelerates the numerical calculation. With $\rho
_{0}^{\left( g\right) }$, the density matrix to the first order is acquired
through replacing $\mathcal{L}_{0}$ by $\mathcal{L}_{0}^{eff}$ in Eq.~(\ref%
{sm3}) as%
\begin{equation}
\left( \mathcal{L}_{0}^{eff}\mp i\omega \right) \rho _{1}^{\left( \pm
\right) }+\mathcal{L}_{1}\rho _{0}^{\left( g\right) }=0,  \label{sm4}
\end{equation}%
where the electron spin operators in $\mathcal{L}_{1}$ are in the
ground-state subspace.

\begin{figure*}[tbh]
\begin{center}
\includegraphics[width=\linewidth]{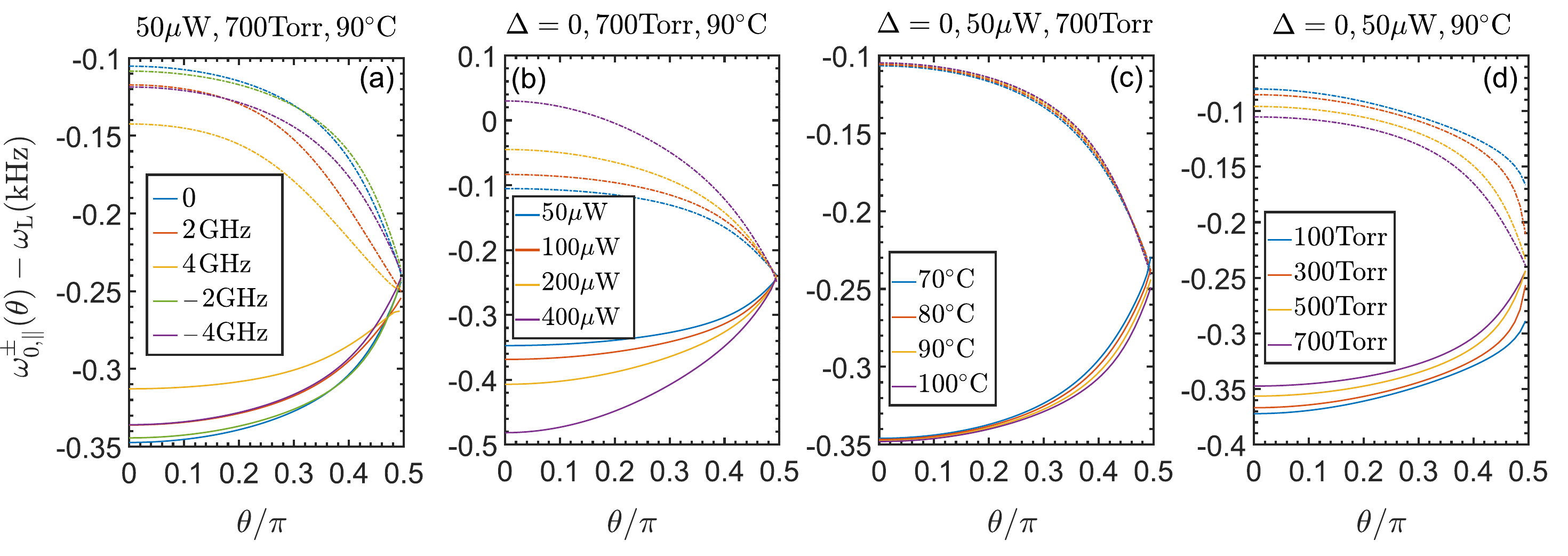}
\end{center}
\caption{Precession frequencies $\protect\omega _{0,\parallel }^{\pm }$ with
respect to the Larmor frequency $\protect\omega _{\mathrm{L}}$ under
different conditions. The external magnetic field $B_0=55\protect\mu$T.}
\label{figSm1}
\end{figure*}

\subsection{Nuclear Zeeman effect on the precession frequency}

The driving Hamiltonian $H_{D}\propto S_{x}\cos \theta -S_{z}\sin \theta $.
Since $\rho _{0}$ has only diagonal terms, in $\left[ S_{z},\rho _{0}\right]
$ only terms $\left\vert F_{1}m\right\rangle _{SS}\left\langle
F_{2}m\right\vert $ with $F_{1}\neq F_{2}$ exists which has large energy $%
\Delta _{S}$ in the superspace. Thus, we can neglect the term $S_{z}\sin
\theta $ in $H_{D}$. We can further apply the rotating-wave approximation
and ignore the term $S_{z}\sin \theta $ in $S_{\parallel }\left( \omega
\right) $ since the mixing rate of $\left\vert Fm\right\rangle
_{SS}\left\langle Fm\right\vert $ and $\left\vert F^{\prime }m\right\rangle
_{SS}\left\langle F^{\prime }m\pm 1\right\vert $ is much smaller that the
Larmor frequency $\omega _{\mathrm{L}}$.

\begin{figure*}[tbh]
\begin{center}
\includegraphics[width=\linewidth]{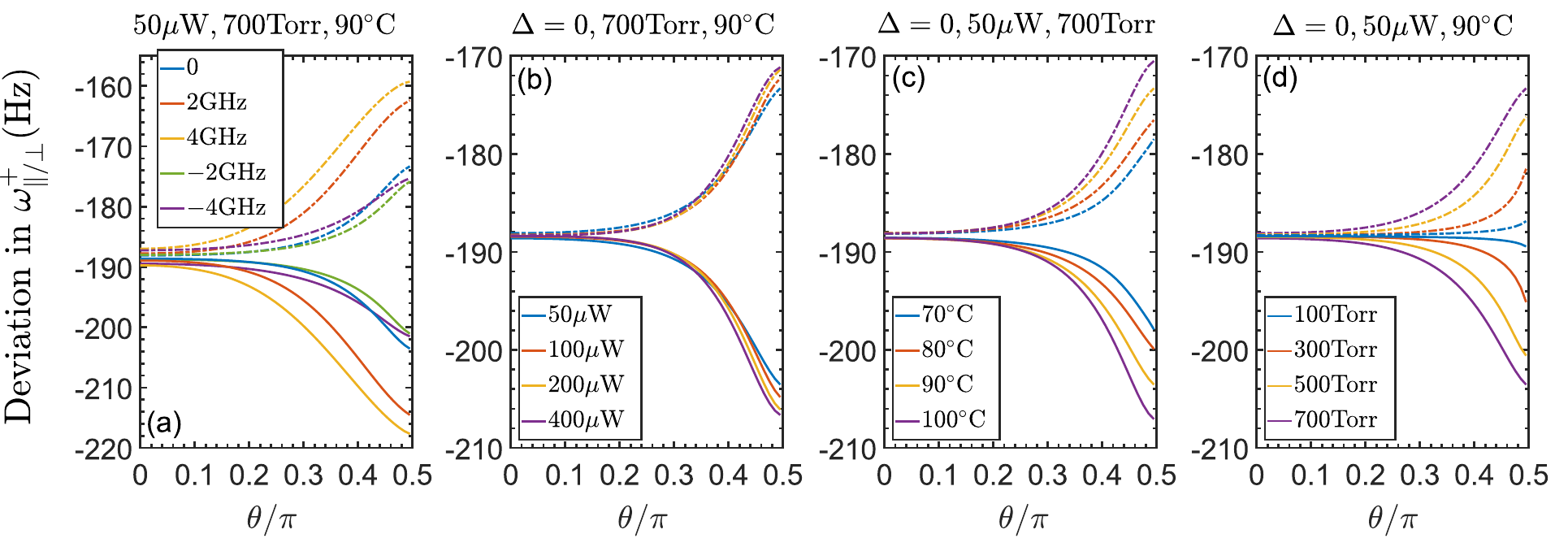}
\end{center}
\caption{Difference of the precession frequencies with and without the NuZ
effect: $\protect\omega _{0,\parallel /\perp }^{+}\left( \text{with NuZ}%
\right) -\protect\omega _{0,\parallel /\perp }^{+}\left( \text{without NuZ}%
\right) $. Here, the solid lines are for the case with parallel probe
lasers, while the dotted-dash lines are for the vertical ones.}
\label{figSm2}
\end{figure*}
Under our experimental conditions, $S_{\parallel /\perp }^{\pm }\left(
\omega \right) =0$ has two solutions around $\pm \omega _{\mathrm{L}}$,
respectively, where in the superscript, \textquotedblleft $+/-$%
\textquotedblright\ represents the $\sigma ^{+}/\sigma ^{-}$ polarization.
Here, we focus on the solution $\omega _{0}$ around $+\omega _{\mathrm{L}}$.
From Eq.~(\ref{sm4}) we have%
\begin{equation}
S_{\parallel }^{\pm }\left( \omega \right) =-2\mathrm{Re}\left[ \mathrm{Tr}%
\left( S_{x}\left( \mathcal{L}_{0}^{eff}\mp i\omega \right) ^{-1}\mathcal{L}%
_{1}\rho _{0}\right) \right] \cos \theta  \label{sm5}
\end{equation}%
and%
\begin{equation}
S_{\perp }^{\pm }\left( \omega \right) =\mp 2\mathrm{Im}\left[ \mathrm{Tr}%
\left( S_{y}\left( \mathcal{L}_{0}^{eff}\mp i\omega \right) ^{-1}\mathcal{L}%
_{1}\rho _{0}\right) \right] \cos \theta .  \label{sm6}
\end{equation}%
The precession frequency for the parallel probe laser case is plotted in
Fig.~\ref{figSm1} for different (a) detunings $\Delta $, (b) pump powers,
(c) temperatures, and (d) densities of nitrogen gas. The external field $%
B_{0}=55\mu $T.

To show quantitatively the NuZ effect on the precession frequency, we plot
the deviation of the precession frequencies with and without considering the
NuZ effect in Fig.~\ref{figSm2} under different conditions as in Fig.~\ref%
{figSm1}. We can see that the NuZ effect lowers the precession frequency by
an amount smaller than the linear NuZ splitting $\omega _{\text{NuZ}}$ ($226$%
Hz) because the $a$-manifold has a frequency $\omega _{\text{L}}-\omega _{%
\text{NuZ}}$ while the $b$-manifold has a frequency $\omega _{\text{L}%
}+\omega _{\text{NuZ}}$. As $\theta $ changes, the deviation $\omega
_{0,\parallel }^{+}\left( \text{with NuZ}\right) -\omega _{0,\parallel
}^{+}\left( \text{without NuZ}\right) $ becomes more negative while $\omega
_{0,\perp }^{+}\left( \text{with NuZ}\right) -\omega _{0,\perp }^{+}\left(
\text{without NuZ}\right) $ gets less negative because of the opposite
contribution from the $b$ manifold (See Sec. ). Here, we only show the
deviation for $\omega _{0,\parallel /\perp }^{+}$ because the contribution
from the NuZ effect to the precession frequency is the same for $\omega
_{0,\parallel /\perp }^{+}$ and $\omega _{0,\parallel /\perp }^{-}$. The
proof is as follows. As shown in Eq. (\ref{sm5}), for $\sigma ^{+}$ pump
laser, $\omega _{0,\parallel }^{+}$ is determined by
\begin{equation}
S_{\parallel }^{+}\left( \omega _{0,\parallel }^{+}\right) =-2\mathrm{Re}%
\left[ \mathrm{Tr}\left( S_{x}\left( \mathcal{L}_{0}^{eff}\left(
B_{0}\right) -i\omega _{0,\parallel }^{+}\right) ^{-1}\mathcal{L}_{1}\rho
_{0}\right) \right] \cos \theta =0.
\end{equation}%
For $\sigma ^{-}$ pump laser, which is equivalent to inverting the magnetic
field $B_{0}$ while keeping the pump laser's helicity the same, $\omega
_{0,\parallel }^{-}$ is determined by%
\begin{equation}
S_{\parallel }^{-}\left( \omega _{0,\parallel }^{-}\right) =-2\mathrm{Re}%
\left[ \mathrm{Tr}\left( S_{x}\left( \mathcal{L}_{0}^{eff}\left(
-B_{0}\right) +i\omega _{0,\parallel }^{-}\right) ^{-1}\mathcal{L}_{1}\rho
_{0}\right) \right] \cos \theta =0.
\end{equation}%
When the NLZ effect and LS are neglected and only the NuZ effect is
considered,
\begin{eqnarray}
S_{\parallel }^{-}\left( \omega _{0,\parallel }^{-}\right) &=&-2\mathrm{Re}%
\left[ \mathrm{Tr}\left( S_{x}\left( \mathcal{L}_{0}^{eff}\left(
-B_{0}\right) +i\omega _{0,\parallel }^{-}\right) ^{-1}\mathcal{L}_{1}\rho
_{0}\right) ^{\ast }\right] \cos \theta  \notag \\
&=&2\mathrm{Re}\left[ \mathrm{Tr}\left( S_{x}\left( \mathcal{L}%
_{0}^{eff}\left( B_{0}\right) -i\omega _{0,\parallel }^{-}\right) ^{-1}%
\mathcal{L}_{1}\rho _{0}\right) \right] \cos \theta  \notag \\
&=&-S_{\parallel }^{+}\left( \omega _{0,\parallel }^{-}\right) .  \label{8}
\end{eqnarray}%
The case for $\omega _{0,\perp }^{+}$ and $\omega _{0,\perp }^{-}$ can be
proved similarly from Eq. (\ref{sm6}). Therefore, $\omega _{0,\parallel
/\perp }^{+}=\omega _{0,\parallel /\perp }^{-}$ when considering only the
NuZ effect.

\subsection{Asymmetric heading errors}

Under the rotating-wave approximation, in $\mathcal{L}_{1}\rho _{0}$ in Eqs.
(\ref{sm5}) and (\ref{sm6}), only terms $\left\vert a,m\right\rangle
_{SS}\left\langle a,m+1\right\vert $ and $\left\vert b,m+1\right\rangle
_{SS}\left\langle b,m\right\vert $ need to be taken into account. In this
basis, the $B_{0}$-dependent diagonal terms of $\mathcal{L}_{0}^{eff}$ are
in the diagonal terms as(ignoring the small modification of the hyperfine
states resulting from the interaction to the external field $\vec{B}_{0}$)%
\begin{equation}
E\left( a,m+1\right) -E\left( a,m\right) \approx \omega _{\mathrm{L}}-\omega
_{\mathrm{NuZ}}-\left( 2m+1\right) \omega _{\mathrm{rev}},
\end{equation}%
\begin{equation}
E\left( b,m\right) -E\left( b,m+1\right) \approx \omega _{\mathrm{L}}+\omega
_{\mathrm{NuZ}}-\left( 2m+1\right) \omega _{\mathrm{rev}}.
\end{equation}%
As a result, $S_{\parallel /\perp }^{+}\left( \omega \right) $ is a function
of $\alpha _{m}^{a,b}\left[ \omega _{\mathrm{L}}-\sigma ^{a,b}\omega _{%
\mathrm{NuZ}}-\left( 2m+1\right) \omega _{\mathrm{rev}}-\omega \right] $,
where $\alpha _{m}^{a,b}$ is a coefficient dependent on the manifold $a$, $b$%
, and the magnetic number $m$, but independent of $\omega $ and $B_{0}$, $%
\sigma ^{a}=1$ and $\sigma ^{b}=-1$. For $\sigma ^{-}$ polarization, or
equivalently for negative $B_{0}$, similarly, in $\mathcal{L}_{1}\rho _{0}$,
only terms $\left\vert a,m\right\rangle _{SS}\left\langle a,m+1\right\vert $
and $\left\vert b,m+1\right\rangle _{SS}\left\langle b,m\right\vert $ need
to be considered and $S_{\parallel /\perp }^{-}\left( \omega \right) $ is
the same/opposite function of $\alpha _{m}^{a,b}\left[ -\omega _{\mathrm{L}%
}+\sigma ^{a,b}\omega _{\mathrm{NuZ}}-\left( 2m+1\right) \omega _{\mathrm{rev%
}}+\omega \right] $. Therefore, the solutions $\omega _{0,\shortparallel
/\perp }^{\pm }$ to the equations $S_{\parallel /\perp }^{+}\left( \omega
_{0,\shortparallel /\perp }^{+}\right) =0$ and $S_{\parallel /\perp
}^{-}\left( \omega _{0,\shortparallel /\perp }^{-}\right) =0$ fulfill%
\begin{equation}
\omega _{0,\parallel /\perp }^{+}+\omega _{0,\parallel /\perp }^{-}=2\omega
_{\mathrm{L}}
\end{equation}%
if the NuZ splitting $\omega _{\mathrm{NuZ}}$ is ignored, i.e., the heading
errors for $\sigma ^{+}$ and $\sigma ^{-}$ polarizations are symmetric:
\begin{equation}
\omega _{0,\parallel /\perp }^{+}\left( \theta _{1}\right) -\omega
_{0,\parallel /\perp }^{+}\left( \theta _{2}\right) =\omega _{0,\parallel
/\perp }^{-}\left( \theta _{2}\right) -\omega _{0,\parallel /\perp
}^{-}\left( \theta _{1}\right) .  \label{sm7}
\end{equation}%
However, the existence of $\omega _{\mathrm{NuZ}}$ does not only shifts $%
\left( \omega _{0,\parallel /\perp }^{+}+\omega _{0,\parallel /\perp
}^{-}\right) /2-\omega _{\mathrm{L}}$ from zero, but breaks the symmetry (%
\ref{sm7}), as shown in Figs.~(\ref{fig2}) and (\ref{fig3}). As shown in the
last subsection, the NuZ effect contributes the same to the precession
frequencies $\omega _{0,\parallel /\perp }^{+}$ and $\omega _{0,\parallel
/\perp }^{-}$, which is $\theta $-dependent, so the heading errors including
all three sources are not symmetric for $\sigma ^{+}$ and $\sigma ^{-}$
polarizations.

The precession frequencies $\omega _{0,\parallel }$ for the parallel probe
laser case and the corresponding heading errors are shown in Fig.~\ref%
{figSm3}\ for different external fields $B_{0}$ and pumping powers, while
other parameters are the same as in Fig.~2(a). When the power of
the pump laser is lower or the external magnetic field is smaller, the
heading errors for both $\sigma ^{+}$ and $\sigma ^{-}$ polarizations are
smaller. The former is because the population changes in the Zeeman states
with smaller Rabi frequencies become smaller as the tilted angle $\theta $
varies, while the latter is because the NLZ and NuZ induced frequency
deviations become smaller. In all these cases, the heading errors for $%
\sigma ^{+}$ and $\sigma ^{-}$ polarizations are asymmetric: with $\sigma
^{+}$-polarized pumps, the heading error is smaller. Same as Fig.~\ref{fig2}%
, here we only compare the theoretical results with the experimental data on
the heading errors, because the real magnetic fields slightly deviate from $%
30$, $50$, $70\mu \mathrm{T}$. These differences can not be determined
accurately in our experiment, but they are small so that the NLZ effect and
the NuZ effect induced by them can be ignored. Therefore, the heading errors
$\omega _{0}\left( \theta \right) -\omega _{0}\left( 0\right) $ are the same
as in the $B_{0}=30$, $50$, $70\mu \mathrm{T}$ cases. The relatively large
deviation between the theoretical and experimental results Figs.~\ref{figSm2}%
(b) and (f) is resulting from the remnant magnetization in the magnetic
shield, which has a more obvious effect for smaller field $B_{0}$.

\begin{figure*}[tbh]
\begin{center}
\includegraphics[width=\linewidth]{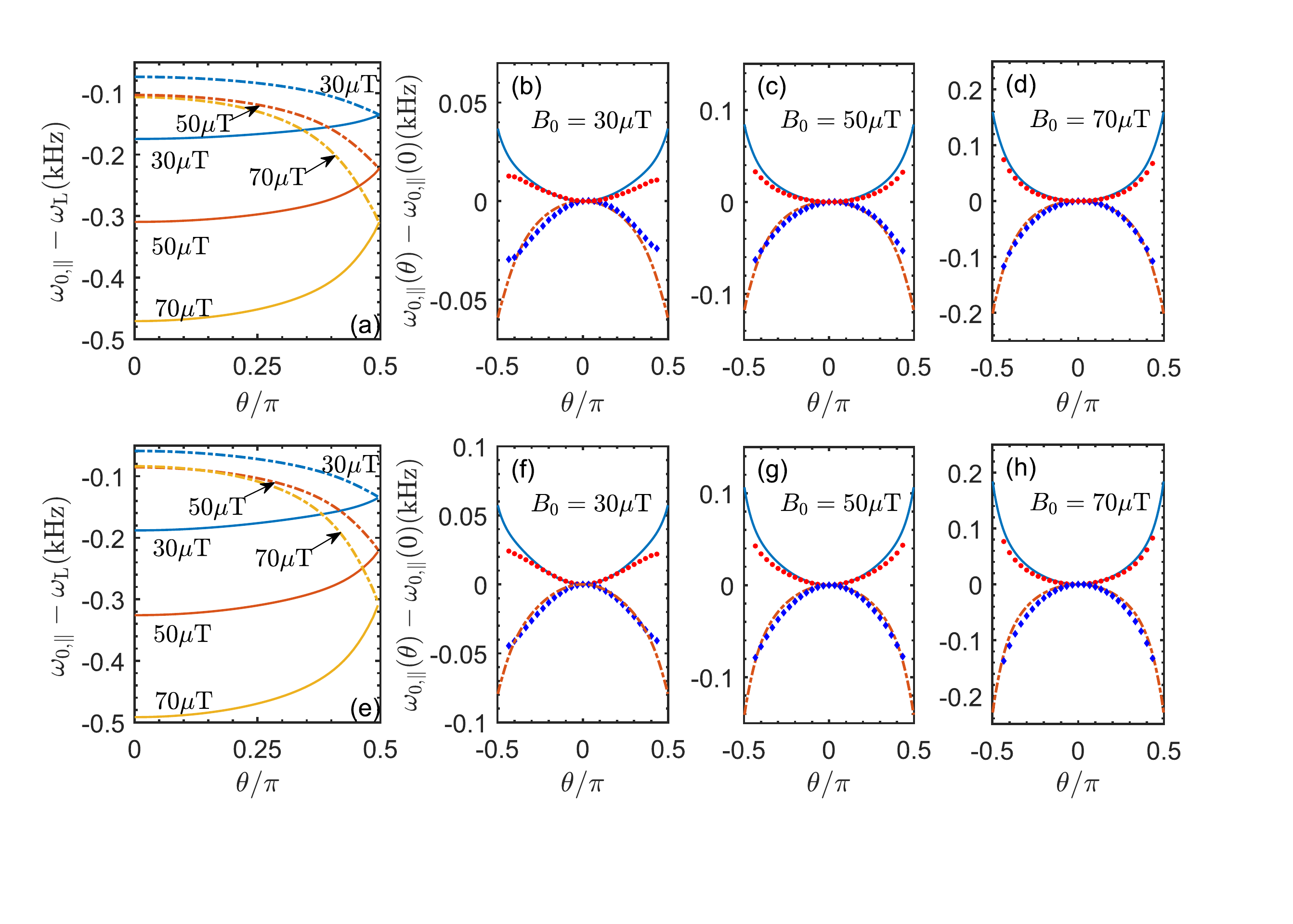}
\end{center}
\caption{(a) Precession frequencies $\protect\omega _{0,\parallel }^{\pm }$
with respect to the Larmor frequency $\protect\omega _{\mathrm{L}}$ and
heading errors $\protect\omega _{0}\left( \protect\theta \right) -\protect%
\omega _{0}\left( 0\right) $ (b)-(d) for different $B_{0}$ with the probe
laser parallel to the RF field. Experimental data are plotted in red dots ($%
\protect\sigma ^{+}$ polarization) and blue diamonds ($\protect\sigma ^{-}$
polarization). The power of the pump laser is $50\protect\mu $W, and its
detuning $\Delta =0$. (e)-(h) Same as (a)-(d), but with a larger pump power
of $90\protect\mu $W$.$}
\label{figSm3}
\end{figure*}

\begin{figure*}[tbh]
\begin{center}
\includegraphics[width=\linewidth]{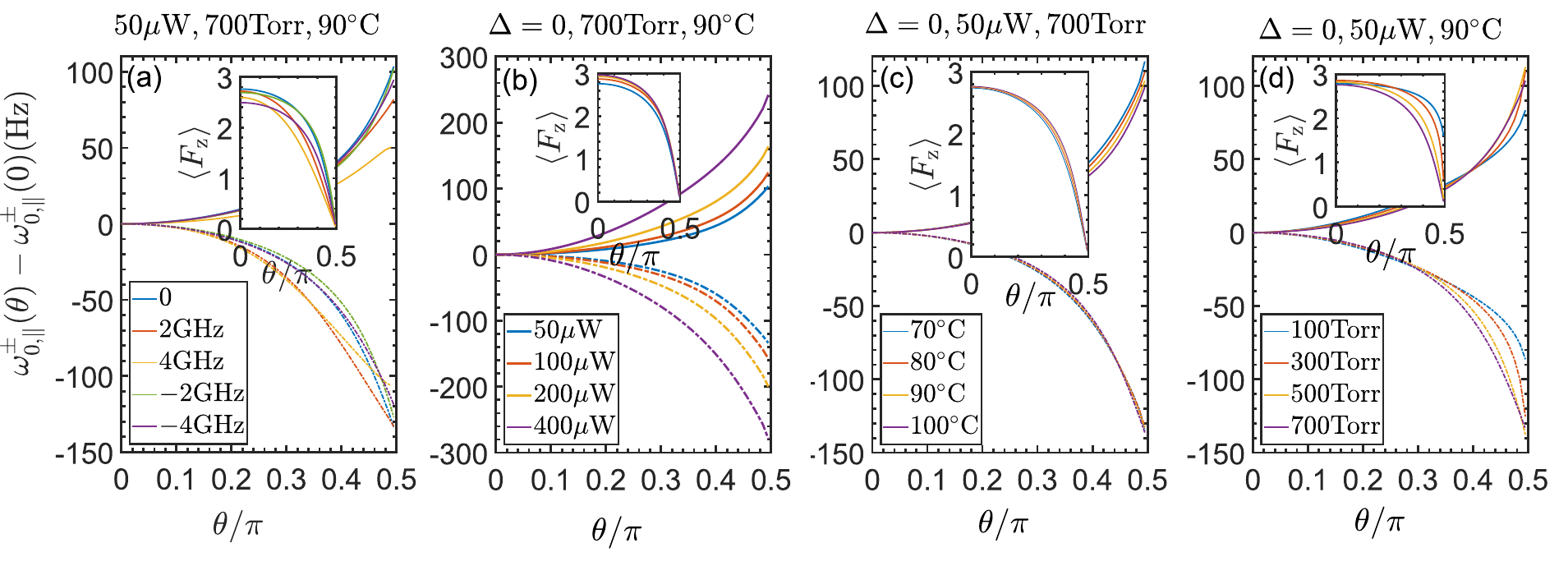}
\end{center}
\caption{Heading errors $\protect\omega _{0,\parallel }^{\pm }\left( \protect%
\theta \right) -\protect\omega _{0,\parallel }^{\pm }\left( 0\right) $ under
various conditions. The inset shows the mean value of the total spin's
angular moment in the $z$ direction $\left\langle F_{z}\right\rangle $ for
the $\protect\sigma ^{+}$-polarized pump. For the $\protect\sigma ^{-}$
case, $\left\langle F_{z}\right\rangle $ is the opposite.}
\label{figSm4}
\end{figure*}

The heading errors in a broad parameter regime for $B_{0}=55\mu \mathrm{T}$
are plotted in Fig.~\ref{figSm4}, which also shows asymmetry for the $\sigma
^{+}$ and $\sigma ^{-}$ polarizations. To give a physical insight of the
heading error under different conditions, we show the average value $%
\left\langle F_{z}\right\rangle $ in the insets, where $F_{z}=S_{z}+I_{z}$
is the total spin's angular moment along the $z$ direction. The mean value $%
\left\langle F_{z}\right\rangle $ reveals the connection between the heading
errors and the atomic populations in the Zeeman levels: the heading error
increases monotonously as $\left\langle F_{z}\right\rangle $ changes faster
as the tilted angle $\theta $ varies. For instance, as shown in Fig.~\ref%
{figSm4}(c), as the temperature increases, the change of the polarization $%
\left\langle F_{z}\right\rangle $\ becomes smaller since the hotter cell has
less nitrogen gas. As a result, the heading error gets smaller.

\section{Reduction of heading errors using two pump lasers with opposite
helicities}

In geophysical surveys to detect magnetic anomalies of various types such as
searching for mineral deposits or locating lost objects, the relative angle
between the object and the magnetometer is not well-known and it varies with
time, leading to accuracy degradation for high-accuracy magnetometers. For
instance, with the parameters shown in Fig. 4(a), to achieve the accuracy of
$1$pT around $\theta =60^{\circ }$, one needs to acquire the angle with an
uncertainty of $\left( 1.56\times 10^{-3}\right) ^{\circ }$ (blue solid
curve), which is almost impossible to reach. Hence, the heading error
results in accuracy degradation and needs to be reduced.

In this section, we study the compensation of the heading errors by summing
up the measured precession frequencies from atomic vapor cells pumped by
lasers with opposite helicities.

\subsection{Both probe lasers parallel or perpendicular to the AC driving
fields}

When the probe lasers are parallel or perpendicular to the AC driving fields
in both cells, it has been proved in the last section (see Eq. (\ref{sm7}))
that the heading errors induced by the NLZ effect and LS can be canceled by
averaging the output from magnetometers pumped by opposite-helicity lasers.
However, the NuZ effect breaks the symmetry of the heading errors in these
two atomic cells, giving the monotonously increasing heading errors as the
tilted angle $\theta $ increases, as shown in Figs. 2(a) and 2(b).

\begin{figure*}[tbh]
\begin{center}
\includegraphics[width=\linewidth]{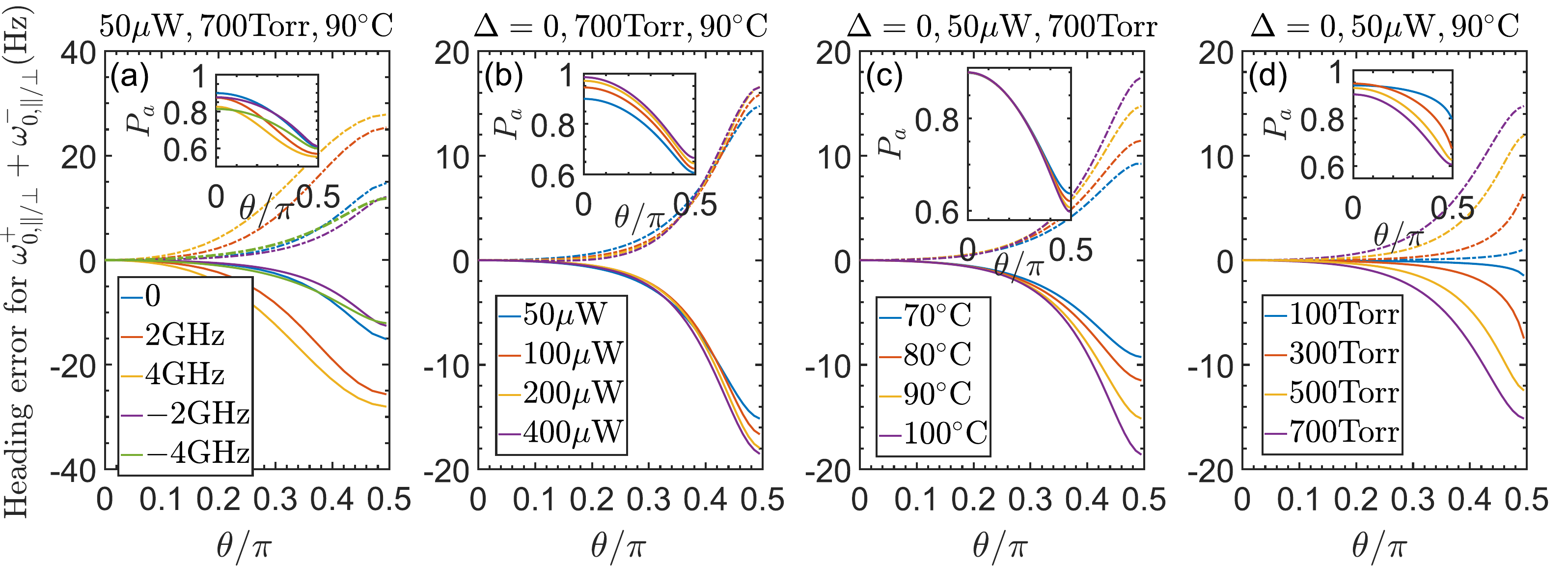}
\end{center}
\caption{Heading errors in the average of two magnetometers pumped by
opposite-helicity lasers and probe by lasers in the same directions. The
external field $B_{0}=55\protect\mu $T. The insets show the atomic
populations in the $a$-manifold.}
\label{figSm5}
\end{figure*}

Since the asymmetry of the heading errors comes from the opposite
contributions of the NuZ splitting $\omega _{\text{NuZ}}$ to the $a$ and $b$
manifolds, increasing the populations or reducing the change of the
populations in the $a$-manifold can reduce the heading error in the averaged
precession frequencies from the two cells. In Fig.~\ref{figSm5}, the heading
errors $\left[ \omega _{0,\parallel /\perp }^{+}\left( \theta \right)
+\omega _{0,\parallel /\perp }^{-}\left( \theta \right) -\omega
_{0,\parallel /\perp }^{+}\left( 0\right) -\omega _{0,\parallel /\perp
}^{-}\left( 0\right) \right] /2$ are shown for $B_{0}=55\mu $T while other
parameters such as the laser' detuning $\Delta $ and power, the temperature,
and the density of the nitrogen gas are varied. The populations in the $a$
manifold $P_{a}=$Tr$\rho _{0}^{\left( a\right) }$ are shown in the
corresponding insets. We see from Fig.~\ref{figSm5}(a) that the heading
error is smaller when the detuning $\Delta $ is around $0$ where the
transition between the $b$-manifold and the excited states is resonant with
the pump laser and the population in the $b$ manifold is suppressed. In Fig.~%
\ref{figSm5}(b), the heading errors do not change much as the power of the
pump laser varies, since lowering the input power decreases not only the
population and but its change in the $a$-manifold. However, the case in (c)
and (d) is different: when lowing the temperature or the density of the
buffer gas, the population in the $a$ manifold increases while its change
decreases, and thus the heading error is reduced. Note that this
uncompensated heading error in Fig.~\ref{figSm5} results from the NuZ
effect, so it does not only show the deviation of the heading errors when
the NuZ is not considered, but gives the improvement one can acquire by
choosing the $\sigma ^{+}$-polarized ($\sigma ^{-}$-polarized) pump laser
other than the $\sigma ^{-}$-polarized ($\sigma ^{+}$-polarized) one with
the parallel (vertical) probe laser.

\subsection{One probe laser parallel and another perpendicular to the AC
driving fields}

Tuning the parameters of the two cells with the probe lasers both parallel
or perpendicular to the driving fields can weaken the NuZ-effect on the
heading errors but can not compensate for it. For this, we propose to employ
one probe laser parallel and another perpendicular to the driving fields in
this two-cell scheme. This can be understood simply as the following: we
approximate $S_{\parallel /\perp }\left( \omega \right) \approx -g_{e}\mu _{%
\mathrm{B}}B_{1}\left[ S^{a}\left( \omega \right) \pm S^{b}\left( \omega
\right) \right] \cos \theta $ with the plus sign for $S_{\parallel }\left(
\omega \right) $ and minus sign for $S_{\perp }\left( \omega \right) $, where%
\begin{equation}
S^{a}\left( \omega \right) =\mathrm{Re}\left[ \mathrm{Tr}\left( S_{+}\left(
\mathcal{L}_{0}^{eff}-i\omega \right) ^{-1}\mathcal{L}_{1-}\rho _{0}^{\left(
a\right) }\right) \right] ,
\end{equation}%
\begin{equation}
S^{b}\left( \omega \right) =\mathrm{Re}\left[ \mathrm{Tr}\left( S_{-}\left(
\mathcal{L}_{0}^{eff}-i\omega \right) ^{-1}\mathcal{L}_{1+}\rho _{0}^{\left(
b\right) }\right) \right] ,
\end{equation}%
for $\sigma ^{+}$\ polarization, while%
\begin{equation}
S^{a}\left( \omega \right) =\mathrm{Re}\left[ \mathrm{Tr}\left( S_{+}\left(
\mathcal{L}_{0}^{eff}+i\omega \right) ^{-1}\mathcal{L}_{1-}\rho _{0}^{\left(
a\right) }\right) \right] ,
\end{equation}%
\begin{equation}
S^{b}\left( \omega \right) =\mathrm{Re}\left[ \mathrm{Tr}\left( S_{-}\left(
\mathcal{L}_{0}^{eff}+i\omega \right) ^{-1}\mathcal{L}_{1+}\rho _{0}^{\left(
b\right) }\right) \right] ,
\end{equation}%
for $\sigma ^{-}$\ polarization (in the treatment here, we invert $\vec{B}%
_{0}$ instead of the healicity of the pump). Here, $\mathcal{L}_{1\pm }\rho
=-i\left[ S_{\pm },\rho \right] $ and $\rho _{0}^{\left( a/b\right) }$ is
the density matrix projected in the $a/b$-manifold: $\rho _{0}^{\left(
F=a,b\right) }=\sum_{mm^{\prime }}\left\vert Fm\right\rangle
_{SS}\left\langle Fm\right\vert \rho _{0}\left\vert Fm^{\prime
}\right\rangle _{SS}\left\langle Fm^{\prime }\right\vert $. Consequently, in
the parallel case, the precession frequency $\omega _{0,\parallel }^{\pm }$
for parallel probe lasers is determined by $S^{a}\left( \omega _{0,\parallel
}^{\pm }\right) +S^{b}\left( \omega _{0,\parallel }^{\pm }\right) =0$ while
the precession frequency $\omega _{0,\perp }^{\pm }$ for the perpendicular
case is determined by $S^{a}\left( \omega _{0,\perp }^{\pm }\right)
-S^{b}\left( \omega _{0,\perp }^{\pm }\right) =0$. Since the contribution
from the $a$-manifold is lager, we expand $S^{a/b}\left( \omega \right) $
around $\omega _{a}^{\pm }$ as%
\begin{equation}
S^{a}\left( \omega \right) \approx C_{a}^{\pm }\left( \omega -\omega
_{a}^{\pm }\right)
\end{equation}%
and%
\begin{equation}
S^{b}\left( \omega \right) \approx C_{b}^{\pm }\left( \omega -\omega
_{a}^{\pm }\right) +D^{\pm },
\end{equation}%
where $C_{a}^{\pm }$, $C_{a}^{\pm }$, and $D^{\pm }$\ are constants with the
plus sign in the superscript for $B_{0}>0$ and minus sign for $B_{0}<0$.
When the NLZ effect and the LS are neglected, similar to Eq. (\ref{sm8})\ we
have%
\begin{equation*}
\omega _{a}^{+}=\omega _{a}^{-}=\omega _{\mathrm{L}}-\omega _{\mathrm{NuZ}},
\end{equation*}%
$C_{a}^{+}=-C_{a}^{-}$, and $D_{a}^{+}=-D_{a}^{-}$. Therefore, the solutions
are%
\begin{equation}
\omega _{0,\parallel }^{\pm }=\omega _{\mathrm{L}}-\omega _{\mathrm{NuZ}}-%
\frac{D^{\pm }}{C_{a}^{\pm }+C_{b}^{\pm }}
\end{equation}%
and%
\begin{equation}
\omega _{0,\perp }^{\pm }=\omega _{\mathrm{L}}-\omega _{\mathrm{NuZ}}+\frac{%
D^{\pm }}{C_{a}^{\pm }-C_{b}^{\pm }}.
\end{equation}%
In the limit $\left\vert C_{a}^{\pm }\right\vert \gg \left\vert C_{b}^{\pm
}\right\vert $, which is the case in our parameter regime, the NuZ-induced
heading errors in $\omega _{0,\parallel }^{\pm }+\omega _{0,\perp }^{\mp }$
are canceled.

\begin{figure*}[tbh]
\begin{center}
\includegraphics[width=\linewidth]{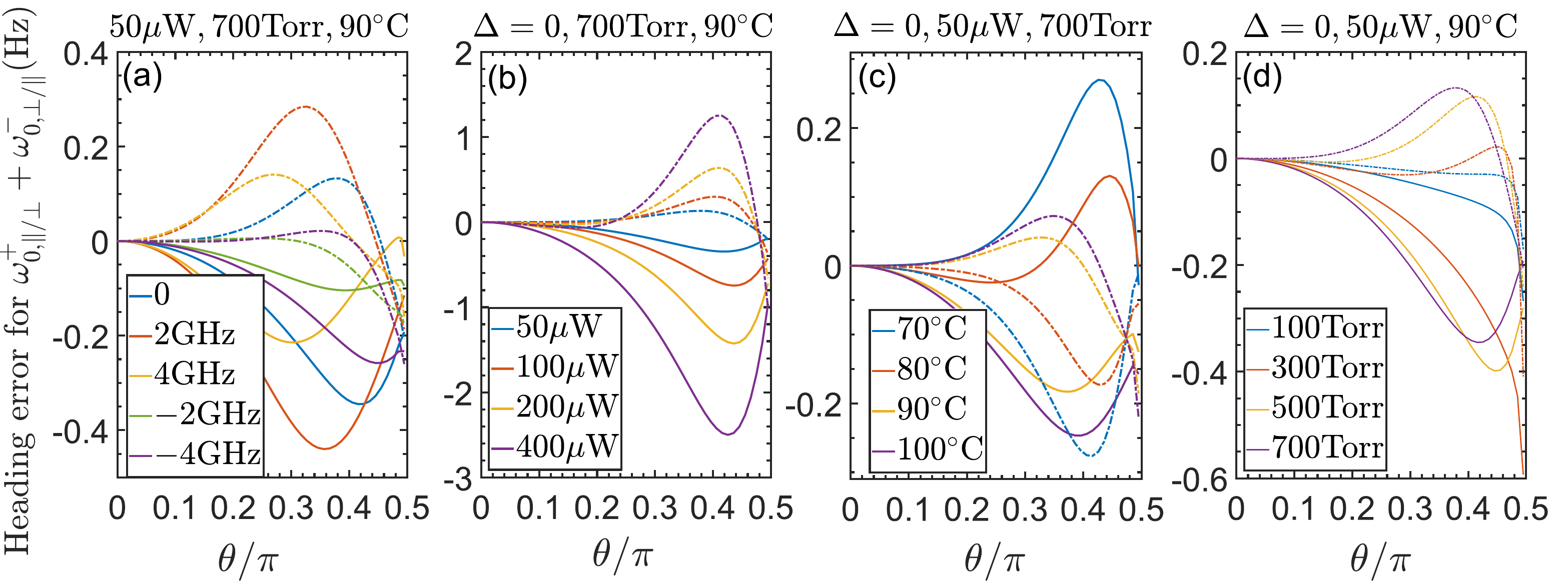}
\end{center}
\caption{Same as Fig.~\protect\ref{figSm4}, but for averaging cells probed
by lasers in different directions (one parallel to the RF driving field
while another perpendicular to it). Here, solid lines are for $\protect%
\omega _{0,\parallel }^{+}+\protect\omega _{0,\perp }^{-}$\ and dotted-dash
lines are for $\protect\omega _{0,\perp }^{+}\left( \protect\theta \right) +%
\protect\omega _{0,\parallel }^{-}\left( \protect\theta \right) $. }
\label{figSm6}
\end{figure*}

With the same parameters as in Fig.~\ref{figSm5}, we plot the heading errors
$\left[ \omega _{0,\parallel /\perp }^{+}\left( \theta \right) +\omega
_{0,\perp /\parallel }^{-}\left( \theta \right) -\omega _{0,\parallel /\perp
}^{+}\left( 0\right) -\omega _{0,\perp /\parallel }^{-}\left( 0\right) %
\right] /2$ in Fig.~\ref{figSm6}. Because of the large cancellation of the
NuZ effect, the heading errors are well suppressed in a wide parameter
regime.

\section{Heading errors compensated by an auxiliary field}

We have shown in Fig.~4 that the measured precession frequency has a smaller
angular dependence by using an auxiliary field. For instance, with an
auxiliary field of $15$nT (purple solid curve in Fig. 4(a)), to achieve the
accuracy of $1$pT around $\theta =60^{\circ }$, the required uncertainty in
the angle $\theta $ becomes $\left( 2.67\times 10^{-2}\right) ^{\circ }$. In
the application, the strength and sometimes the direction of the magnetic
field in the circumstance where the magnetometer works are roughly known.
Therefore, we can calibrate the magnetometer in the lab with an auxiliary
field, considering a 2\% difference of the magnetic field strength. Within
this difference, we have shown our auxiliary-field method is robust.

In this section, we give an insight into our heading-error-compensation
method by employing a small auxiliary magnetic field, and compare it with
some other methods in the literature. At the large relaxation regime, the
atomic system's density matrix can be approximated by the spin-temperature
distribution \cite{PhysRevA.58.1412}
\begin{equation}
\rho =\exp \left( \beta F_{z}\right) /\text{Tr}\left[ \exp \left( \beta
F_{z}\right) \right] ,
\end{equation}%
where the parameter $\beta $ is connected to the electron's polarization as
\begin{equation}
\exp \left( \beta \right) =\left( 1+2\left\langle S_{z}\right\rangle \right)
/\left( 1-2\left\langle S_{z}\right\rangle \right) .
\end{equation}%
Applying the linear response theory and summing up incoherently the
transition frequencies between two adjacent states using the weight in $%
\left[ S_{x},\rho \right] $, one can acquire the precession frequency $%
\omega _{0}$. Note from Eq. (4), the polarization $\left\langle
S_{z}\right\rangle $ for $\sigma ^{+}$-polarized pump\ can be approximated
as
\begin{eqnarray}
\left\langle S_{z}\right\rangle  &=&\frac{R_{op}\left[ \left( \cos \theta
+1\right) ^{2}-\left( \cos \theta -1\right) ^{2}\right] }{2R_{op}\left[
\left( \cos \theta +1\right) ^{2}+\left( \cos \theta -1\right) ^{2}\right]
+2\Gamma _{rel}}  \notag \\
&=&\frac{2R_{op}\cos \theta }{R_{op}\left[ \left( \cos \theta +1\right)
^{2}+\left( \cos \theta -1\right) ^{2}\right] +2\Gamma _{rel}}.
\end{eqnarray}%
where $R_{op}$ is the optical pumping rate for a $\sigma ^{+}$-polarized
pump laser that can be obtained by adiabatically eliminating the excited
states, and $\Gamma _{rel}$ is the relaxation rate. Therefore, for small $%
\theta $, one can expand $\cos \theta $ around $1$ and obtain the angular
dependence of $\omega _{0}$ as $-\cos \theta $ for $\sigma ^{+}$-polarized
pump laser and $\cos \theta $ for the $\sigma ^{-}$-polarized pump laser
(note that the polarization $\left\langle S_{z}\right\rangle $ is inverted
for $\sigma ^{-}$-polarized pump). With the small auxiliary field, the total
precession frequency is shown in Eq. (7) where the cosine function can
cancel the angular dependence in $\omega _{0}\left( \theta ,B_{a}=0\right) $
by properly choosing $B_{a}$. This is why the compensation is excellent in
Fig.4(b) for small $\theta $. For large $\theta $, the compensation is
getting worse since higher orders terms of $\cos \theta $ become more
important. However, mathematically, one can always find an amplitude $B_{a}$
to make the derivative $\partial _{\theta }\omega _{0}\left( \theta
,B_{a}\right) =\partial _{\theta }\omega _{0}\left( \theta ,B_{a}=0\right) -%
\mathrm{sign}\left( \sigma ^{\pm }\right) \mu _{\mathrm{eff}}B_{\mathrm{a}%
}\sin \theta $ vanish at some angle $\theta _{0}$. Then the heading error is
flattened within some angular interval around $\theta _{0}$, as shown in
Fig. 4(c).

Our method is a compensation way to reduce the heading error. Following the
analysis above, we can obtain a lengthy expression of the precession
frequency. But we have theoretically simplified the model by, for instance,
ignoring the incoherent coupling between each adjacent transitions, which
may result in deviations from the exact result, and in the setup, there
might be remnant magnetic field parallel or antiparallel to the sensor's
direction that can also be compensated by the auxiliary field, so the best
way in practical use is to measure the heading error under controllable
conditions first, then determine the auxiliary field according to the
measured data.

In the literature, apart from the compensation methods using two
magnetometers pumped by opposite-helicity lasers \cite%
{doi:10.1063/1.1663412,scholtes2012light,PhysRevA.99.013420}, some other
methods are proposed to reduce the heading error. For example, the
cancelation of NLZ effect with LS \cite%
{PhysRevA.79.023406,PhysRevA.82.023417}, the spin-locking method using a RF
field \cite{PhysRevLett.120.033202}, and the synchronous optical pumping
with double-modulation of the lasers \cite{PhysRevA.75.051407}. However,
they have all concentrated on the NLZ effect and the LS, but neglected the
NuZ effect. Thus when the atomic population in the $b$ manifold (or the $a$
manifold if the pump laser is resonant with it) increases, which is usually
the case when the sensor is more tilted (note that it is not the case in the
pulsed-pump magnetometer where the populations in each manifold remain the
same), those methods are less effective \cite{lee2021heading}. In contrast,
our method by using the auxiliary field can still flatten the heading error
at large $\theta $ by tuning the auxiliary field $B_{a}$, which helps
increase the accuracy of magnetometers working around a certain angle.

%\bibliographystyle{unsrtnat}
%\bibliographystyle{plainnat}
%\bibliography{ref}
%\bibliography{HeadingError1.bbl}

\end{document}